\documentclass[aps,prc,superscriptaddress,nofootinbib,showpacs,floatfix,singlecolumn,longbibliography]{revtex4-2}
\usepackage{url}
\usepackage{cancel}
\usepackage[colorlinks,linkcolor=blue,citecolor=blue,filecolor=black,urlcolor=blue]{hyperref}
\usepackage{epsfig,graphics}
\usepackage{graphicx}% Include figure files
\usepackage{dcolumn}% Align table columns on decimal point
\usepackage{bm}% bold math
\usepackage[usenames]{color}
\usepackage{amssymb}
\usepackage{amsmath}
\usepackage{multirow}
\usepackage{float}
\usepackage{harpoon}
\usepackage{MnSymbol}
\usepackage{appendix}
\usepackage{color}
\usepackage{hyperref}
\usepackage{cleveref}

\newcommand{\sqrtsnn}{\mbox{$\sqrt{s_{\mathrm{NN}}}$}}
\newcommand{\pT} {p_{\mathrm{T}}}
\newcommand{\lr}[1]{\left\langle #1\right\rangle}
\newcommand{\llrr}[1]{\left\llangle #1\right\rrangle}

\newcommand{\nall}{N_{\mathrm{hadron}}}
\newcommand{\npart}{N_{\mathrm{part}}}
\newcommand{\nqp}{N_{\mathrm{quark}}}

\usepackage{CJK}
\hfuzz=\maxdimen
\begin{document}
\title{Probing triaxial deformation of atomic nuclei in high-energy heavy ion collisions}
\newcommand{\sbu}{Department of Chemistry, Stony Brook University, Stony Brook, NY 11794, USA}
\newcommand{\bnl}{Physics Department, Brookhaven National Laboratory, Upton, NY 11976, USA}
\author{Jiangyong Jia}\email[Correspond to\ ]{jiangyong.jia@stonybrook.edu}\affiliation{\sbu}\affiliation{\bnl}
\date{\today}
\begin{abstract}
Most atomic nuclei are deformed with a quadrupole shape described by its overall strength $\beta_2$ and triaxiality $\gamma$. The deformation can be accessed in high-energy heavy-ion collisions by measuring the collective flow response of the produced quark-gluon plasma to the eccentricity $\varepsilon_2$ and the density gradient $d_{\perp}$ in the initial state. Using an analytical estimate and a Glauber model, I show that the variances, $\langle\varepsilon_2^2\rangle$ or $\langle(\delta d_{\perp}/d_{\perp})^2\rangle$, and skewnesses, $\langle\varepsilon_2^2\delta d_{\perp}/d_{\perp}\rangle$ or $\langle(\delta d_{\perp}/d_{\perp})^3\rangle$, have a simple analytical form of $a'+b'\beta_2^2$ and $a'+(b'+c'\cos(3\gamma))\beta_2^3$, respectively. From these, I constructed several normalized skewnesses to isolate the $\gamma$ dependence from that of $\beta_2$, and show that the correlations between a normalized skewness and a variance can constrain simultaneously the $\beta_2$ and $\gamma$. Assuming a linear relation with elliptic flow $v_2$ and mean-transverse momentum $[p_{\mathrm{T}}]$ of final-state particles, $v_2\propto \varepsilon_2$ and $\delta[p_{\mathrm{T}}]/[p_{\mathrm{T}}] \propto \delta d_{\perp}/d_{\perp}$, similar conclusions are also expected for the variances and skewnesses of $v_2$ and $[p_{\mathrm{T}}]$, i.e. $a+b\beta_2^2$ for $\langle v_2^2\rangle$ and $\langle (\delta [p_{\mathrm{T}}]/[p_{\mathrm{T}}])^2\rangle$ and $a+(b+c\cos(3\gamma))\beta_2^3$ for $\langle v_2^2\delta [p_{\mathrm{T}}]/[p_{\mathrm{T}}]\rangle$ or $\langle(\delta [p_{\mathrm{T}}]/[p_{\mathrm{T}}])^3\rangle$.  These findings motivate a dedicated system scan of high-energy heavy-ion collisions at RHIC and LHC to measure triaxiality of atomic nuclei: one first determines the coefficients $b$ and $c$ by collisions of isobaric near prolate nuclei, $\cos(3\gamma)\approx1$, and near oblate nuclei, $\cos(3\gamma)\approx-1$, with known $\beta_2$ values, followed by collisions of other species of interest with similar mass number. The $(\beta_2,\gamma)$ values for this species can be inferred directly from the measured variance and skewness observables from these collisions. The results demonstrate the unique opportunities offered by high-energy collisions as a tool to perform interdisciplinary nuclear physics studies.
\end{abstract}
\maketitle
%\tableofcontents \clearpage
\section{Introduction}\label{sec:1}

Most atomic nuclei in their ground state are deformed from a well-defined spherical shape. Nuclear deformation arises due to short-range strong nuclear force among nucleons themselves, and depending on the proton and neutron number, the minima in the total energy of the system can be found for spherical, ellipsoidal, octuple and hexadecapole shapes~\cite{Heyde2011,Togashi:2016yzs,Heyde:2016sop,Frauendorf:2017ryj,Zhou:2016ujx}. Information about nuclear deformation is primarily extracted from spectroscopic measurements and models of reduced transition probability $B(En)$ between low-lying rotational states, which involves nuclear experiments with energy per nucleon less than few tens of MeVs. Recently, the prospects of probing the nuclear deformation at much higher beam energy, energy per nucleon exceeding hundreds of GeVs, by taking advantage of the hydrodynamic flow behavior of large number of produced final-state particles, have been discussed~\cite{Heinz:2004ir,Filip:2009zz,Shou:2014eya,Goldschmidt:2015kpa,Giacalone:2017dud,Giacalone:2018apa,Giacalone:2021uhj,Giacalone:2021udy,Jia:2021wbq,Jia:2021tzt,Bally:2021qys}, and evidence from several experiments has been observed~\cite{Adamczyk:2015obl,Acharya:2018ihu,Sirunyan:2019wqp,ATLAS:2019dct,jjia}. 

\begin{figure}[!t]
\begin{center}
\hspace*{0.4cm}\includegraphics[width=0.95\linewidth]{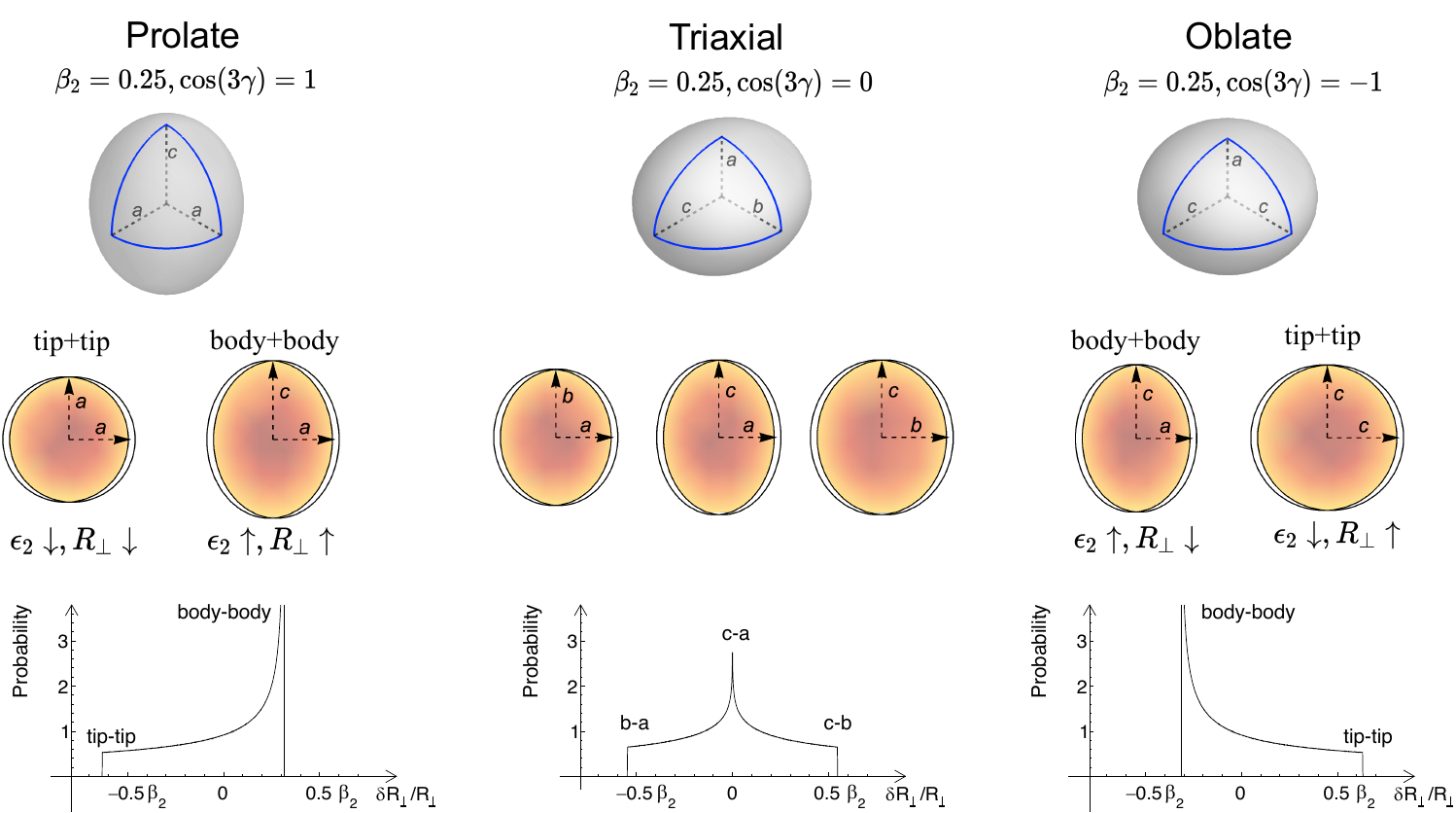}
\end{center}
\caption{\label{fig:1} The cartoon of a nucleus with quadrupole deformation $\beta_2=0.25$ (top row) with prolate (left), rigid triaxial (middle), and oblate (right) shape, the overlap containing the quark-gluon plasma in the ultracentral collisions (middle row) and distributions of the transverse size, $\delta R_{\perp}/R_{\perp}=-\delta d_{\perp}/d_{\perp}$, derived from Eq.~\eqref{eq:11} (bottom row). The distributions in the bottom are given in units of $\beta_2$.}
\end{figure}

The shape of a nucleus, including only the dominant quadrupole component, is often described by a nuclear density profile of the Woods-Saxon form,
\begin{align}\label{eq:1}
\rho(r,\theta,\phi)=\frac{\rho_0}{1+e^{\left[r-R(\theta,\phi)/a\right]}},\;R(\theta,\phi) = R_0\left(1+\beta_2 [\cos \gamma Y_{2,0}(\theta,\phi)+ \sin\gamma Y_{2,2}(\theta,\phi)]\right),
\end{align}
where the nuclear surface $R(\theta,\phi)$ is expanded into real form spherical harmonics $Y_{2,m}$ in the intrinsic frame. The positive number $\beta_2$ describes the overall quadrupole deformation, and the triaxiality parameter $\gamma$ controls the relative order of the three radii $r_a,r_b,r_c$ of the nucleus in the intrinsic frame. It has the range $0\leq\gamma\leq\pi/3$, with $\gamma=0$, $\gamma=\pi/3$, and $\gamma=\pi/6$ corresponding, respectively, to prolate ($r_a=r_b<r_c)$, oblate ($r_a<r_b=r_c$) or rigid triaxiality ($r_a<r_b<r_c$ and $2r_b=r_a+r_c$), see top row of Fig.~\ref{fig:1} for an illustration. Most nuclei have axially symmetric prolate or oblate shapes, and triaxiality is a rather elusive signature in nuclear structure physics. The triaxial degree of freedom is related to a number of interesting phenomena including the $\gamma$-band~\cite{bohr}, chirality~\cite{Frauendorf:1997mux} and wobbling motion~\cite{PRL.86.5866,RevModPhys.73.463}, but the extraction of $\gamma$ value has significant experimental and theoretical uncertainties. An interesting question is if and how triaxiality may manifest itself in other fields of nuclear physics.

High-energy heavy-ion collisions at RHIC and the LHC, especially head-on collisions with nearly zero impact parameter (ultracentral collisions or UCC), provide a new way to image the shape of the nucleus. The large amount of energy deposited in these collisions leads to the formation of a hot and dense quark-gluon plasma (QGP)~\cite{Busza:2018rrf} in the overlap region, whose shape and size are strongly correlated with nuclear deformation as illustrated by the second row of Fig.~\ref{fig:1}. The transverse area $S_{\perp}$ (or size $R_{\perp}$) and eccentricity of the overlap can be quantified by
\begin{align}\label{eq:2}
 S_{\perp}\equiv  \pi R^2_{\perp}= \pi\sqrt{\lr{x^2}\lr{y^2}}\;,\;\;\;{\bf \epsilon_2} \equiv \varepsilon_2e^{i2\Phi_2} = - \frac{\lr{r_{\perp}^2 e^{i2\phi}}}{\lr{r_{\perp}^2}},
\end{align}
where the average is over nucleons in the transverse plane $(x,y)=(r_{\perp},\phi)$ in the rotated center-of-mass frame such that $x$ ($y$) corresponds to the minor (major) axis of the ellipsoid. Within the liquid-drop model with a sharp surface, the variances of $\varepsilon_2$ and $R_{\perp}$ over many head-on collisions are directly related to the $\beta_2$: $\lr{\varepsilon_2^2}=\frac{3}{2\pi}\beta_2^2$, $\lr{(\delta R_{\perp}/R_{\perp})^2} = -\frac{1}{16\pi} \beta_2^2$, where $\delta R_{\perp}/R_{\perp} \equiv (R_{\perp}-\lr{R_{\perp}})/\lr{R_{\perp}}$ denotes the event-by-event fluctuations relative to the average. Driven by the large pressure gradient forces and subsequent hydrodynamic collective expansion, the initial shape and size information is transferred into azimuthal and radial flow of final-state hadrons~\cite{Heinz:2013wva}. Specifically, the particle momentum spectra in each collision event can be parametrized as $\frac{d^2N}{\pT d\pT d\phi} = N(\pT) \left[1+2v_2(\pT) \cos 2(\phi-\Psi)\right]$ in $\phi$ and transverse momentum $\pT$. The magnitude of the radial flow, characterized by the slope of the particle spectrum $N(\pT)$ or the average $[\pT]$, is positively correlated with the gradient of nucleon density or inverse transverse size $d_{\perp}$
\begin{align}\label{eq:2b}
d_{\perp} =\sqrt{\npart/S_{\perp}},
\end{align}
in the overlap region~\cite{Bozek:2012fw,Schenke:2020uqq}, with $\npart$ being the number of participating nucleons. This is because $d_{\perp}\propto 1/R_{\perp}$ is proportional to the pressure gradient and therefore is expected to be correlated with $[\pT]$.  Similarly, the amplitude and orientation of elliptic flow, characterized by $V_2=v_2e^{i2\Psi}$, is directly related to ${\bf \epsilon_2}=\varepsilon_2e^{i2\Phi}$. In fact, detailed hydrodynamic model simulations~\cite{Niemi:2015qia,Schenke:2020uqq} show good linear relations, for events with fixed $\npart$.
\begin{align}\label{eq:3}
v_2=k_2\varepsilon_2,\;\;\; \frac{\delta [\pT]}{[\pT]} = k_0\frac{\delta d_{\perp}}{d_{\perp}} = -k_0\frac{\delta R_{\perp}}{R_{\perp}} =-k_0\frac{1}{2}\frac{\delta S_{\perp}}{S_{\perp}}\;.
\end{align}
The response coefficients $k_2$ and $k_0$ capture the transport properties of the QGP and they have been constrained theoretically~\cite{Teaney:2012ke,Bernhard:2016tnd,Bernhard:2019bmu,Nijs:2020ors}.

As indicated clearly in the second row of Fig.~\ref{fig:1}, in ultracentral collisions of prolate nuclei, the shape of overlap falls in between ``body-body'' and ``tip-tip'' configurations with the long-axis perpendicular or parallel to the beam, respectively. The body-body collisions have large $\varepsilon_2$ and larger size $R_{\perp}$ and therefore smaller $d_{\perp}$, while the tip-tip collisions have near-zero $\varepsilon_2$ and larger $d_{\perp}$, i.e. the correlation of $\varepsilon_2$ and $d_{\perp}$ is  negative $\lr{\varepsilon_2^2\delta d_{\perp}}<0$~\cite{Giacalone:2019pca}. In contrast, the covariance of $\varepsilon_2$ and $d_{\perp}$ is expected to be positive for collisions of oblate nuclei, and zero for collisions of rigid triaxial nuclei~\cite{Jia:2021wbq}. Eq.~\eqref{eq:3} would then imply that $\lr{v_2^2\delta[\pT]}<0$, $>0$ and $=0$ for collisions of prolate, oblate and rigid triaxial nuclei, respectively. In fact, one find that both $\lr{\varepsilon_2^2\delta d_{\perp}}$ and $\lr{v_2^2\delta[\pT]}$ are dominated by a $\cos(3\gamma)$ dependence in the ultracentral collisions, not surprising given the three-fold symmetry of nuclear shape in the $\gamma$ angle.

Another interesting aspect of the deformation in heavy ion collisions, not discussed yet in the literature, concerns the nature of the event-by-event fluctuations of $R_{\perp}$ or $d_{\perp}$ in the presence of deformation and how they influence the $[\pT]$ fluctuations. As shown in the bottom row of Fig.~\ref{fig:1}, the probability for various overlap configurations are not equal. In collisions of rigid triaxial nuclei, the shape of the overlap in the transverse plane falls in between three configurations for the two axes of the ellipse: ``$r_b-r_a$'', ``$r_c-r_a$'' and ``$r_c-r_b$''. The combination ``$r_c-r_a$'' has the largest probability, and as the nucleus becomes more prolate (oblate), the middle branch merges with the right (left) branch and  the distribution becomes more asymmetric. This gives rise to a nonvanishing skewness $\lr{(\delta d_{\perp})^3}\sim -\lr{(\delta R_{\perp})^3}$, and the sign of $\lr{(\delta d_{\perp})^3}$ is expected to be opposite to that of $\lr{\varepsilon_2^2\delta d_{\perp}}$. Indeed, one finds that $\lr{(\delta d_{\perp})^3}$ contains a large $\cos (3\gamma)$ term, which is expected to drive a similar term for the skewness $\lr{(\delta [\pT])^3}$ in the final state. Therefore, I have identified two three-particle correlation observables, $\lr{v_2^2\delta[\pT]}$ and $\lr{(\delta [\pT])^3}$, to probe nuclear triaxiality in heavy ion collisions. The $\beta_2$ value on the other hand can be constrained from two-particle correlation observables $\lr{v_2^2}$ and $\lr{(\delta [\pT])^2}$.

Several experimental studies of nuclear deformation in heavy ion collisions have been carried at RHIC~\cite{Adamczyk:2015obl,Giacalone:2021udy} and the LHC~\cite{Acharya:2018ihu,Sirunyan:2019wqp,ATLAS:2019dct}, focusing mostly on the relation between $\beta_2$ and $v_2$ in the UCC. However, the most striking evidence is provided by the recent measurement of $\lr{v_2^2\delta[\pT]}$ and $\lr{(\delta [\pT])^3}$ in $^{197}$Au+$^{197}$Au and $^{238}$U+$^{238}$U collisions at RHIC~\cite{jjia}. The large prolate deformation of $^{238}$U yields a large negative contribution to $\lr{v_2^2\delta[\pT])}$ and a large positive contribution to $\lr{(\delta [\pT])^3}$, consistent with the picture in Fig.~\ref{fig:1} discussed above. A few model studies on the feasibility of constraining triaxiality in heavy ion collisions appeared recently~\cite{Jia:2021wbq,Jia:2021tzt,Bally:2021qys}. In light of these measurements and model work, I aim to clarify, via a Monte-Carlo Glauber model and a transport model, the influence of deformation on the cumulants of $\varepsilon_2$ and $[\pT]$. Remarkably, one finds that the $\beta_2$ and $\gamma$ dependencies of these observables follow very simple parametric functional forms. In particular, one finds that $\lr{\varepsilon_2^2}$ and $\lr{\delta (d_{\perp})^3}$ can be well described by a function of the $a'+b'\beta_2^2$ form, while $\lr{\varepsilon_2^2\delta d_{\perp}}$ and $\lr{\delta (d_{\perp})^3}$ by a function of the $a'+(b'+c'\cos(3\gamma))\beta_2^3$ form, with $b'$ and $c'$ nearly independent of the size of the collision systems. This finding provides a motivation for a collision system scan of nuclei at ultrarelativistic energies with similar $\beta_2$ but different $\gamma$ values, which may provide additional insight on the question of shape evolution and shape coexistence~\cite{Heyde2011} in low-energy nuclear structure physics.

\section{Simple analytical estimate}\label{sec:2}
I first predict the analytical form for the $(\beta_2,\gamma)$ dependencies using a simple heuristic argument. For small deformation $\beta_2$, the values of $d_{\perp}$ and ${\bm \epsilon}_2$ in a given event are expected to have the following form:
\begin{align}\label{eq:4}
\frac{\delta d_{\perp}}{d_{\perp}} = \delta_d + p_0(\Omega_1,\Omega_2,\gamma)\beta_2 + \mathcal{O}(\beta_2^2)\;,\;{\bm \epsilon}_2 = {\bm \epsilon}_0 + {\bm p}_{2}(\Omega_1,\Omega_2,\gamma)\beta_2 + \mathcal{O}(\beta_2^2),
\end{align}
where the scalar $\delta_d$ and vector ${\bm \epsilon}_0=\varepsilon_0e^{i2\Phi_{2;0}}$ are values for spherical nuclei, which in UCC collisions are dominated by random fluctuations of nucleons positions but in noncentral collisions are also affected by the impact-parameter-dependent average shape of the overlap. The $p_0$ and ${\bm p}_{2}$ are phase-space factors controlled by the Euler angles $\Omega=\phi\theta\psi$ of the two nuclei; they also contain the $\gamma$ parameter. For example, in collision of prolate nuclei (see left of the middle row of Fig~\ref{fig:1}), $|p_0|$ and $|{\bm p}_{2}|$ are largest for the ``body-body'' orientation and smallest for the ``tip-tip'' orientation. Since the fluctuation of $\delta_d$ (${\bm \epsilon}_0$) is uncorrelated with $p_0$ (${\bm p}_{2}$), an average over collisions with different Euler angles is expect to give the following expression for the variances
\begin{align}\label{eq:5}
C_{\mathrm{d}}\{2\}\equiv\lr{\left(\frac{\delta d_{\perp}}{d_{\perp}}\right)^2} = \lr{\delta_d^2} + \lr{p_0(\Omega_1,\Omega_2,\gamma)^2}\beta_2^2\;,\;
c_{2,\epsilon}\{2\}\equiv\lr{\varepsilon_2^2} = \lr{\varepsilon_0^2} +  \lr{{\bm p}_{2}(\Omega_1,\Omega_2,\gamma){\bm p}_{2}^*(\Omega_1,\Omega_2,\gamma)}\beta_2^2\;.
\end{align}
The $\lr{p_0^2}$ and $\lr{{\bm p}_{2}{\bm p}_{2}^*}$ are constants obtained by averaging over $\Omega_1$ and $\Omega_2$. This argument can be generalized to higher-order cumulants. For example, the skewness and kurtosis of $p(d_{\perp})$ and kurtosis of ${\bm \epsilon}_2$ can be written as,
\begin{align}\nonumber
C_{\mathrm{d}}\{3\}&\equiv\lr{\left(\frac{\delta d_{\perp}}{d_{\perp}}\right)^3} = \lr{\delta_d^3} + \lr{p_0^3}\beta_2^3\;,\\\nonumber
C_{\mathrm{d}}\{4\}&\equiv\lr{\left(\frac{\delta d_{\perp}}{d_{\perp}}\right)^4}-3\lr{\left(\frac{\delta d_{\perp}}{d_{\perp}}\right)^2}^2 = \lr{\delta_d^4}-3\lr{\delta_d^2}^2 + (\lr{p_0^4}-3\lr{p_0^2}^2)\beta_2^4\;\\\label{eq:6}
c_{2,\epsilon}\{4\}&\equiv\lr{\varepsilon_2^4}-2\lr{\varepsilon_2^2}^2 = \lr{\varepsilon_0^4}-2\lr{\varepsilon_0^2}^2 + \left(\lr{{\bm p}_{2}^2{\bm p}_{2}^{*2}}-2\lr{{\bm p}_{2}{\bm p}_{2}^{*}}^2\right)\beta_2^4\;,
\end{align}
where I use the fact that $\lr{{\bm p}_{2}^n{\bm p}_{2}^{*m}}=0$ for $n\neq m$ due to the invariance under a global rotation. I shall skip the straightforward expression for higher-order cumulant of ${\bm \epsilon}_2$. Another interesting example is mixed-skewness $\lr{\varepsilon_2^2\frac{\delta d_{\perp}}{d_{\perp}}}$, a good estimator for $\lr{v_2^2\frac{\delta [\pT]}{[\pT]}}$,
\begin{align}\label{eq:7}
\lr{\varepsilon_2^2 \frac{\delta d_{\perp}}{d_{\perp}}} = \lr{\varepsilon_0^2\delta_d} +\lr{p_0{\bm p}_{2}{\bm p}_{2}^*}\beta_2^3\;.
\end{align}
Note that in noncentral collisions, the cross-term like $\lr{p_0({\bm p}_{2}{\bm \epsilon}_{0}^*+{\bm p}_{2}^*{\bm \epsilon}_{0})}\beta_2^2$ term may not vanish due to possible alignment between ${\bm \epsilon}_{0}$ and ${\bm p}_{2}$.

This argument can be generalized to simultaneous presence of octuple or hexadecapole deformations for which additional axial symmetric components are added to the nuclear surface in Eq.~\eqref{eq:1},
\begin{align}\label{eq:1b}
R(\theta,\phi) = R_0\left(1+\beta_2 [\cos \gamma Y_{2,0}(\theta,\phi)+ \sin\gamma Y_{2,2}(\theta,\phi)] + \beta_3 Y_{3,0}(\theta,\phi)+\beta_4 Y_{4,0}(\theta,\phi)\right)\;,
\end{align}
as well as to the higher-order eccentricities of the overlap region in the transverse plane, defined as ${\bm \epsilon}_n\equiv\varepsilon_ne^{in\Phi_n} = - \lr{r_{\perp}^n e^{in\phi}}/\lr{r_{\perp}^n}$. In this case, the leading order expression for $\delta d_{\perp}$ and eccentricity are $\delta d_{\perp}/d_{\perp} = \delta_d + \sum_{m=2}^{4}p_{0;m}\beta_m$ and ${\bm \epsilon}_n \approx {\bm \epsilon}_{n;0} + \sum_{m=2}^{4} {\bm p}_{n;m}(\Omega_1,\Omega_2)\beta_m$, respectively. The variances have the following more general form 
\begin{align}\label{eq:10}
\lr{\left(\frac{\delta d_{\perp}}{d_{\perp}}\right)^2} \approx \lr{\delta_d^2} + \sum_{m,m'}\lr{p_{0;m}p_{0;m'}}\beta_m\beta_{m'}\;,\;
\lr{\varepsilon_n^2}\approx \lr{\varepsilon_{n;0}^2} + \sum_{m,m'} \lr{{\bm p}_{n;m}{\bm p}^*_{n;m'}}\beta_m\beta_{m'}\;.
\end{align}
The off-diagonal coefficients $\lr{p_{0;m'}p_{0;m'}}_{m\neq m'}$ and  $\lr{{\bm p}_{n;m}{\bm p}^*_{n;m'}}_{m\neq m'}$ may not vanish especially in the non-central collisions. These mixing contributions have been observed in my previous study of $\lr{\varepsilon_n^2}$~\cite{Jia:2021wbq}, and are expected to influence all other cumulants discussed above. I leave this interesting topic to a future study.

For a more quantitative estimation, I consider the liquid-drop model where the nucleon density distribution has a sharp surface. I limit the discussion to head-on collisions with nearly maximum overlap, i.e. the two nuclei not only have zero impact parameter, but are also aligned $\Omega_1=\Omega_2$ to ensure the overlap region contains all the nucleons $\npart=2A$. In this case it is easy to show (see Ref.~\cite{Jia:2021tzt} and Appendix~\ref{sec:app1})
\begin{align}\label{eq:11}
\frac{\delta d_{\perp}}{d_{\perp}}=\sqrt{\frac{5}{16 \pi}} \beta_{2}\left(\cos \gamma D_{0,0}^{2}+\frac{\sin \gamma}{\sqrt{2}}\left[D_{0,2}^{2}+D_{0,-2}^{2}\right]\right)\;,\;{\bm \epsilon}_{2}=-\sqrt{\frac{15}{2 \pi}} \beta_{2}\left(\cos \gamma D_{2,0}^{2}+\frac{\sin \gamma}{\sqrt{2}}\left[D_{2,2}^{2}+D_{2,-2}^{2}\right]\right)\;,
\end{align}
where the $D^{l}_{m,m'}(\Omega)$ is the Wigner matrix. From this, one obtain directly the probability density distributions of $\delta d_{\perp}/d_{\perp}$ shown in the bottom row of Fig.~\ref{fig:1} (the distribution for the prolate case was previously derived in a different context~\cite{Alhassid:2014fca}). From these, one can easily integrate to obtain the expression for cumulants of any order, e.g.:
\begin{align}\nonumber
\left\langle\left(\frac{\delta d_{\perp}}{d_{\perp}}\right)^{2}\right\rangle&=\beta_{2}^{2} \frac{5}{16 \pi} \int\left(\sum_{m} \alpha_{2, m} D_{0, m}^{2}\right)^{2} \frac{d \Omega}{8 \pi^{2}}=\frac{1}{16 \pi} \beta_{2}^{2}\;,\;\;  \alpha_{2,0}\equiv \cos\gamma,\;\alpha_{2,\pm2}\equiv \frac{\sin\gamma}{\sqrt{2}}, \\\nonumber
\left\langle\left(\frac{\delta d_{\perp}}{d_{\perp}}\right)^{3}\right\rangle&=\beta_{2}^{3}\left(\frac{5}{16 \pi}\right)^{3/2} \int\left(\sum_{m} \alpha_{2, m} D_{0, m}^{2}\right)^{3} \frac{d \Omega}{8 \pi^{2}}=\frac{\sqrt{5}}{224 \pi^{3/2}} \cos (3 \gamma) \beta_{2}^{3} \\\label{eq:12}
\left\langle\varepsilon_{2}^{2} \frac{\delta d_{\perp}}{d_{\perp}}\right\rangle&=\beta_{2}^{3} \frac{15}{2 \pi} \sqrt{\frac{5}{16\pi}} \int\left(\sum_{m} \alpha_{2, m} D_{2, m}^{2}\right)\left(\sum_{m} \alpha_{2, m} D_{2, m}^{2}\right)^{*}\left(\sum_{m} \alpha_{2, m} D_{0, m}^{2}\right) \frac{d \Omega}{8 \pi^{2}}=-\frac{3 \sqrt{5}}{28 \pi^{3 / 2}} \cos (3 \gamma) \beta_{2}^{3}\;.
\end{align}
The results for several cumulants of interest are listed in the Table~\ref{tab:1}~\footnote{The expression for $5^{\mathrm{th}}$- and $6^{\mathrm{th}}$-order cumulants of $d_{\perp}$ are $C_d\{5\}=-\frac{15\sqrt{5}}{9856\pi^{5/2}}\cos(3\gamma)\beta_2^5$ and $C_d\{6\}=\frac{15}{7007\times512\pi^3}(113-90\cos(6\gamma))\beta_2^6$.}. If one uses the transverse nucleon density $\npart/S_{\perp}=d_{\perp}^2$ as the estimator as done in Ref.~\cite{Schenke:2020uqq}, the $n^{\mathrm{th}}$-order cumulant would be larger by $2^{n}$. The values for appropriately normalized cumulants are also given to the lower-right side of the observable.

The skewness and kurtosis of $d_{\perp}$ are conventionally normalized by the variance,
\begin{align}\label{eq:13}
S_d=\frac{C_d\{3\}}{C_d\{2\}^{3/2}}\;,\;K_d=\frac{C_d\{4\}}{C_d\{2\}^{2}}\;. 
\end{align}
The four and six-order cumulants of ${\bm \epsilon}_2$ are defined by $\mathrm{nc}_{2,\epsilon}\{4\} = (\lr{\varepsilon_2^4}-2\lr{\varepsilon_2^2}^2)/\lr{\varepsilon_2^2}^2$ and $\mathrm{nc}_{2,\epsilon}\{6\} =\left(\lr{\varepsilon_2^6}-9\lr{\varepsilon_2^4}\lr{\varepsilon_2^2}+12\lr{\varepsilon_2^2}^3\right)/(4\lr{\varepsilon_2^2}^3)$, respectively. The normalization of $\lr{\varepsilon_{2}^{2}\delta d_{\perp}/d_{\perp}}$ is defined in two different ways,
\begin{align}\label{eq:14a}
&\rho_{\mathrm{orig}}(\varepsilon_2^2,\delta d_{\perp}/d_{\perp}) = \frac{\lr{\varepsilon_{2}^{2}\delta d_{\perp}/d_{\perp}}}{\sqrt{(\lr{\varepsilon_2^4}-\lr{\varepsilon_2^2}^2)\lr{\left( d_{\perp}/d_{\perp}\right)^2}}}\;,\;\rho(\varepsilon_2^2,\delta d_{\perp}/d_{\perp}) = \frac{\lr{\varepsilon_{2}^{2}\delta d_{\perp}/d_{\perp}}}{\lr{\varepsilon_2^2}\sqrt{\lr{\left( d_{\perp}/d_{\perp}\right)^2}}}\;.
\end{align}
The $\rho_{\mathrm{orig}}$ is the original definition known as the Pearson correlation coefficient~\cite{Bozek:2016yoj,Schenke:2020uqq}. The term involving $\varepsilon_2$ in its denominator can be expressed as,
\begin{align}\label{eq:14b}
\lr{\varepsilon_2^4}-\lr{\varepsilon_2^2}^2 \equiv \lr{\varepsilon_2^2}^2+c_{2,\varepsilon}\{4\} =\lr{\varepsilon_0^4}-\lr{\varepsilon_0^2}^2 +2\lr{\varepsilon_0^2}\lr{{\bm p}_{2}{\bm p}_{2}^{*}}\beta_2^2+\left(\lr{{\bm p}_{2}^2{\bm p}_{2}^{*2}}-\lr{{\bm p}_{2}{\bm p}_{2}^{*}}^2\right)\beta_2^4\;.
\end{align}
This expression unfortunately contains also an annoying $\beta_2^2$ term that mixes nucleon fluctuations with deformation, which becomes dominant in the mid-central and peripheral collisions. The second definition, $\rho$, preferred in this paper, avoid such analytical complication. But for completeness, the values for both are quoted in Table~\ref{tab:1}.

The normalization of four-particle symmetric cumulants between $\varepsilon_{2}$ and $\delta d_{\perp}$ is defined as
\begin{align}\label{eq:16}
\mathrm{nc}(\varepsilon_2^2,\left(\delta d_{\perp}/d_{\perp}\right)^2) = \frac{\lr{\varepsilon_{2}^{2}\left(\delta d_{\perp}/d_{\perp}\right)^2} -\lr{\varepsilon_{2}^{2}}\lr{\left(\delta d_{\perp}/d_{\perp}\right)^2} }{\lr{\varepsilon_{2}^{2}}\lr{\left(\delta d_{\perp}/d_{\perp}\right)^2}}\;.
\end{align}
This correlator should be measurable with a few hundred millions of events in large systems. Lastly I also calculated the three-particle mixed harmonics $\lr{ {\bm \epsilon}_2^2{\bm \epsilon}_4^*}$, the $\beta_2^4$ dependence arises because the ${\bm\epsilon}_4$ has a $\beta_2^2$ dependence~\cite{Jia:2021tzt}. Interestingly, in the presence of only quadrupole deformation, one has $\lr{ {\bm \epsilon}_2^2{\bm \epsilon}_4^*}=\lr{\varepsilon_4^2}=\frac{45}{14\pi^2}\beta_2^4$. To limit the scope of this paper, I shall skip the discussion of these two observables and the fourth- and higher-order cumulants of $\varepsilon_2$.

The results in Table~\ref{tab:1} are obtained with the assumption $\Omega_1=\Omega_2$. In reality, the selection of UCC events naturally encompasses a wider range of rotation angles and also a finite range of $\npart$, therefore I also study a second case which requires zero impact parameter but independent orientation for the two nuclei. Since the contributions of the two nuclei are independent, the additive nature of the cumulants implies that the value of the $n^{\mathrm{th}}$-order cumulant of intensive quantity is reduced by a factor of $2^{n-1}$, i.e a factor two smaller for $C_d\{2\}$ and $\lr{\varepsilon_2^2}$, a factor of four smaller for $C_d\{3\}$, and a factor of eight smaller for $C_d\{4\}$ and $\lr{\varepsilon_2^4}-2\lr{\varepsilon_2^2}^2$ etc. These values are provided in Tab.~\ref{tab:2}. In realistic model study, $\Omega_1$ and $\Omega_2$ are expected to be only partially aligned and the results for these observables are expected to be in between those given in Tab.~\ref{tab:1} and Tab.~\ref{tab:2}.

\begin{table}[!h]
\centering
\begin{tabular}{c|c||c|c||c|c}\hline
  \multicolumn{2}{c||}{\multirow{2}{*}{$\lr{(\delta d_{\perp}/d_{\perp})^2}$}} & \multicolumn{2}{c||}{\multirow{2}{*}{$\lr{(\delta d_{\perp}/d_{\perp})^3}$}} & \multicolumn{2}{c}{\multirow{2}{*}{$\lr{(\delta d_{\perp}/d_{\perp})^4}-3\lr{(\delta d_{\perp}/d_{\perp})^2}^2$}}\\
 \multicolumn{2}{c||}{}&\multicolumn{2}{c||}{}&\multicolumn{2}{c}{}\\\hline
  \multicolumn{2}{c||}{\multirow{2}{*}{$\frac{1}{16\pi}\beta_2^2$}} & \multirow{2}{*}{$\frac{\sqrt{5}}{224 \pi^{3/2}}\cos(3\gamma)\beta_2^3$} & \multirow{2}{*}{$\frac{2\sqrt{5}}{7}\cos(3\gamma)$} &  \multirow{2}{*}{$-\frac{3}{896 \pi^{2}}\beta_2^4$} &\multirow{2}{*}{$-6/7$}\\
 \multicolumn{2}{c||}{}&&&&\\\hline\hline

  \multicolumn{2}{c||}{\multirow{2}{*}{$\lr{\varepsilon_2^2}$}} & \multicolumn{2}{c||}{\multirow{2}{*}{$\lr{\varepsilon_2^4}-2\lr{\varepsilon_2^2}^2$}} & \multicolumn{2}{c}{\multirow{2}{*}{$\left(\lr{\varepsilon_2^6}-9\lr{\varepsilon_2^4}\lr{\varepsilon_2^2}+12\lr{\varepsilon_2^2}^3\right)/4$}}\\
 \multicolumn{2}{c||}{}&\multicolumn{2}{c||}{}&\multicolumn{2}{c}{}\\\hline
  \multicolumn{2}{c||}{\multirow{2}{*}{$\frac{3}{2\pi}\beta_2^2$}} & \multirow{2}{*}{$-\frac{9}{7\pi^2}\beta_2^4$} & \multirow{2}{*}{$-4/7$} & \multirow{2}{*}{$\frac{27(373-25\cos(6\gamma))}{8008\pi^{3}}\beta_2^6$} & \multirow{2}{*}{$\frac{373-25\cos(6\gamma)}{1001}$} \\
 \multicolumn{2}{c||}{}&&&&\\\hline\hline

  \multicolumn{2}{c||}{\multirow{2}{*}{$\lr{\varepsilon_2^2(\delta d_{\perp}/d_{\perp})}$}} & \multicolumn{2}{c||}{\multirow{2}{*}{$\lr{\varepsilon_2^2(\delta d_{\perp}/d_{\perp})^2}-\lr{\varepsilon_2^2}\lr{(\delta d_{\perp}/d_{\perp})^2}$}} & \multicolumn{2}{c}{\multirow{2}{*}{$\lr{ {\bm \epsilon}_2^2{\bm \epsilon}_4^*}$}}\\
 \multicolumn{2}{c||}{}&\multicolumn{2}{c||}{}&\multicolumn{2}{c}{}\\\hline
  \multirow{2}{*}{$-\frac{3 \sqrt{5}}{28\pi^{3/2}} \cos(3\gamma)\beta_2^3$}&\multirow{2}{*}{$-\frac{2\sqrt{5}}{7}\cos(3\gamma)$,$-\sqrt{\frac{20}{21}} \cos(3\gamma)$}& \multirow{2}{*}{$-\frac{3}{112\pi^2}\beta_2^4$}& \multirow{2}{*}{$-1/4$} & \multicolumn{2}{c}{\multirow{2}{*}{$\frac{45}{14\pi^2}\beta_2^4$}}\\
&&&& \multicolumn{2}{c}{}\\\hline
\end{tabular}
\caption{\label{tab:1} The value of various cumulants of $\varepsilon_2$ and $d_{\perp}$, calculated for nucleus with sharp surface by setting $a=0$ in Eq.~\eqref{eq:1}. The two nuclei are placed with zero impact parameter and results are obtained by averaging over common random orientations. For many observables, I also provide the values after normalizing with second-order cumulants, which are listed in the bottom-right half of the cell (In the case of $\lr{\varepsilon_2^2(\delta d_{\perp}/d_{\perp})}$, both values of $\rho$ (the first number) and $\rho_{\mathrm{orig}}$ (the second number) are provided).}
\end{table}

\begin{table}[!h]
\centering
\begin{tabular}{c|c||c|c||c|c}\hline
  \multicolumn{2}{c||}{\multirow{2}{*}{$\lr{(\delta d_{\perp}/d_{\perp})^2}$}} & \multicolumn{2}{c||}{\multirow{2}{*}{$\lr{(\delta d_{\perp}/d_{\perp})^3}$}} & \multicolumn{2}{c}{\multirow{2}{*}{$\lr{(\delta d_{\perp}/d_{\perp})^4}-3\lr{(\delta d_{\perp}/d_{\perp})^2}^2$}}\\
 \multicolumn{2}{c||}{}&\multicolumn{2}{c||}{}&\multicolumn{2}{c}{}\\\hline
  \multicolumn{2}{c||}{\multirow{2}{*}{$\frac{1}{32\pi}\beta_2^2$}} & \multirow{2}{*}{$\frac{\sqrt{5}}{896 \pi^{3/2}}\cos(3\gamma)\beta_2^3$} & \multirow{2}{*}{$\frac{\sqrt{10}}{7}\cos(3\gamma)$} &  \multirow{2}{*}{$-\frac{3}{7168 \pi^{2}}\beta_2^4$} &\multirow{2}{*}{$-3/7$}\\
 \multicolumn{2}{c||}{}&&&&\\\hline\hline

  \multicolumn{2}{c||}{\multirow{2}{*}{$\lr{\varepsilon_2^2}$}} & \multicolumn{2}{c||}{\multirow{2}{*}{$\lr{\varepsilon_2^4}-2\lr{\varepsilon_2^2}^2$}} & \multicolumn{2}{c}{\multirow{2}{*}{$\left(\lr{\varepsilon_2^6}-9\lr{\varepsilon_2^4}\lr{\varepsilon_2^2}+12\lr{\varepsilon_2^2}^3\right)/4$}}\\
 \multicolumn{2}{c||}{}&\multicolumn{2}{c||}{}&\multicolumn{2}{c}{}\\\hline
  \multicolumn{2}{c||}{\multirow{2}{*}{$\frac{3}{4\pi}\beta_2^2$}} & \multirow{2}{*}{$-\frac{9}{56\pi^2}\beta_2^4$} & \multirow{2}{*}{$-2/7$} & \multirow{2}{*}{$\frac{27(373-25\cos(6\gamma))}{32\times 8008\pi^{3}}\beta_2^6$} & \multirow{2}{*}{$\frac{373-25\cos(6\gamma)}{4004}$} \\
 \multicolumn{2}{c||}{}&&&&\\\hline\hline

  \multicolumn{2}{c||}{\multirow{2}{*}{$\lr{\varepsilon_2^2(\delta d_{\perp}/d_{\perp})}$}} & \multicolumn{2}{c||}{\multirow{2}{*}{$\lr{\varepsilon_2^2(\delta d_{\perp}/d_{\perp})^2}-\lr{\varepsilon_2^2}\lr{(\delta d_{\perp}/d_{\perp})^2}$}} & \multicolumn{2}{c}{\multirow{2}{*}{$\lr{ {\bm \epsilon}_2^2{\bm \epsilon}_4^*}$}}\\
 \multicolumn{2}{c||}{}&\multicolumn{2}{c||}{}&\multicolumn{2}{c}{}\\\hline
  \multirow{2}{*}{$-\frac{3 \sqrt{5}}{112\pi^{3/2}} \cos(3\gamma)\beta_2^3$}&\multirow{2}{*}{$-\frac{\sqrt{10}}{7}\cos(3\gamma)$,$-\sqrt{\frac{2}{7}} \cos(3\gamma)$}& \multirow{2}{*}{$-\frac{3}{896\pi^2}\beta_2^4$}& \multirow{2}{*}{$-1/8$} & \multicolumn{2}{c}{\multirow{2}{*}{$\frac{45}{56\pi^2}\beta_2^4$}}\\
&&&& \multicolumn{2}{c}{}\\\hline
\end{tabular}
\caption{\label{tab:2} Same calculation as Table~\ref{tab:1}, except assuming independent random orientations for the two nuclei.}
\end{table}

A few remarks are in order. The skewness $\lr{\varepsilon_2^2(\delta d_{\perp}/d_{\perp})}$ and $\lr{\left(\delta d_{\perp}/d_{\perp}\right)^3}$ show clear sensitivity to triaxiality in the form of a characteristic $\cos(3\gamma)$ dependence, but with opposite sign. Therefore, when the nuclear shape is varied from prolate to oblate, $\lr{\varepsilon_2^2(\delta d_{\perp}/d_{\perp})}$ is expected to change from negative to positive, while $\lr{(\delta d_{\perp}/d_{\perp})^3}$ is expected to change from positive to negative. In particular, the normalized skewness $\rho$ and $S_{\mathrm{d}}$, defined in Eqs.~\eqref{eq:14a} and \eqref{eq:13}, have equal magnitudes, suggesting a comparable sensitivity to the triaxiality. Secondly, all two- and four-particle correlators have no explicit $\gamma$ dependence, while the six-particle eccentricity cumulant contains a small $\cos(6\gamma)$ modulation. An interesting case is the normalized fourth-order cumulant of $\varepsilon_2$, $\mathrm{nc}_{2}\{4\}=\lr{v_2^4}/\lr{v_2^2}^2-2=-2/7$. Assuming linear-response relation $v_2\{2k\}=k_2\varepsilon_2\{2k\}$ and a large $\beta_2$, one expects a large four-particle cumulant signal $v_2\{4\}$, $v_2\{4\}/v_2\{2\} =\varepsilon_2\{4\}/\varepsilon_2\{2\} \equiv (-\mathrm{nc}_{2}\{4\})^{1/4}=0.73$. This naturally explains the much larger $v_2\{4\}$ value in $^{238}$U+$^{238}$U collisions than that in $^{197}$Au+$^{197}$Au collisions due to the large $\beta_2$ for $^{238}$U nucleus~\cite{Adamczyk:2015obl}. 

\section{Model setup}\label{sec:3}
For a more realistic estimation of influence of nuclear deformation, a Monte-Carlo Glauber model~\cite{Miller:2007ri} is used to simulate collisions of $^{238}$U and $^{96}$Zr systems. These systems are chosen because the experimental collision data exist already. The setup of the model and the data used in this analysis are exactly the same as those used in my previous work~~\cite{Jia:2021tzt}. The nucleons are assumed to have a hard-core of 0.4 fm in radii, with a density described by Eq.~\eqref{eq:1}. The nuclear radius $R_0$ and the surface thickness $a$ are chosen to be $R_0=6.81$~fm and $a=0.55$~fm for $^{238}$U and $R_0=5.09$~fm and $a=0.52$~fm for $^{96}$Zr, respectively. The nucleon-nucleon inelastic cross-section is chosen to be $\sigma_{\mathrm{nn}}=42$~mb at $\sqrtsnn=200$ GeV. In each collision event, nucleons are generated in each nucleus at a random impact parameter. Each nucleus is then rotated by randomly generated Euler angles before they are set on a straight line trajectory towards each other along the $z$ direction. From this, the nucleons in the overlap region are identified, which are used to calculate the $\varepsilon_2$ and $d_{\perp}$ defined in Eqs.~\eqref{eq:2} and \eqref{eq:3}, and the results are presented as a function of $\npart$. Most of the study focuses on the influence of quadrupole deformation, but I also performed a limited study on the influence of the observables from octuple and hexadecapole deformations, for which additional axial symmetric component are added to the nuclear surface (see Eq.~\eqref{eq:1b}). A special study is performed to also investigate the presence of multiple shape components, where two or three nonzero values for $\beta_2$, $\beta_3$ and $\beta_4$ are enabled.

It is well known that particle production in nucleus-nucleus collisions scale only approximately with $\npart$. A better scaling can be achieved by considering the constituent quarks as effective degrees-of-freedom for particle production~\cite{Adler:2013aqf,Lacey:2016hqy,Loizides:2016djv,Bozek:2016kpf,Acharya:2018hhy}, which would naturally give rise to slightly different $\varepsilon_2$  and $d_{\perp}$ in each event. Defining centrality with constituent quarks is also expected to change the fluctuations of eccentricity, and provides a way to quantify the centrality smearing effects (also known as volume fluctuations)~\cite{Skokov:2012ds,Zhou:2018fxx,Aaboud:2019sma}. For this purpose, a quark Glauber model from Ref.~\cite{Loizides:2016djv} is used. Three quark constituents are generated for each nucleon according to the ``mod'' configuration~\cite{Mitchell:2016jio}, which ensures that the radial distribution of the three constituents after re-centering follows the proton form factor $\rho_{\mathrm{proton}}(r) = e^{-r/r_0}$ with $r_0=0.234$~fm~\cite{DeForest:1966ycn}. The value of quark-quark cross-section is chosen to be $\sigma_{\mathrm{qq}}=8.2$~mb in order to match the $\sigma_{\mathrm{nn}}$. The $\varepsilon_2$ and $d_{\perp}$ are then calculated from the list of quark participants in the overlap region, and the number of quark participants $\nqp$ is used as an alternative centrality estimator. 

In the presence of large deformation, the total volume of the nucleus increases slightly. Considering the quadrupole deformation only, for the largest value considered, $\beta_2=0.34$, the ratio to the original volume is approximately (exact for sharp surface nucleus) $1+\frac{3}{4\pi}\beta_2^2+\frac{\sqrt{5}}{28\pi^{3/2}}\cos(3\gamma)\beta_2^3=1.021+0.0004\cos(3\gamma)$. To keep the overall volume fixed, it would require less than 1\% decrease of the $R_0$, which is safely ignored in the present study.

The results for each cumulant observable are obtained in four different ways. Taking the variance $\lr{(\delta d_{\perp}/d_{\perp})^2}$ for instance,  $d_{\perp}$ in each event is calculated either from nucleons or quarks in the Glauber model, after which the averaging ``$\lr{}$'' is then performed for events with the same $\npart$ or the same $\nqp$.  The latter can produce different variances due to slightly different volume fluctuations which can be quite important in the UCC region. Each cumulant can be obtained from either nucleons or quarks and then plotted as a function of $\npart$ or $\nqp$.

I also carried out an independent study based on AMPT transport model to understand the conversion from $\varepsilon_2$ and $d_{\perp}$ in the initial overlap to $v_2$ and $[\pT]$ in the final state. Unfortunately, this model is known to have the wrong hydrodynamic response for the radial flow~\cite{Ma:2016fve,Jia:2021wbq}, therefore it is only used to study the parametric dependence of various observables on $(\beta_2,\gamma)$ and compare with the trends in the initial state.  The detail of the model and the study are presented in Appendix~\ref{sec:app0}.

\section{Results}\label{sec:4}
\begin{figure}[h]
\begin{center}
\includegraphics[width=1\linewidth]{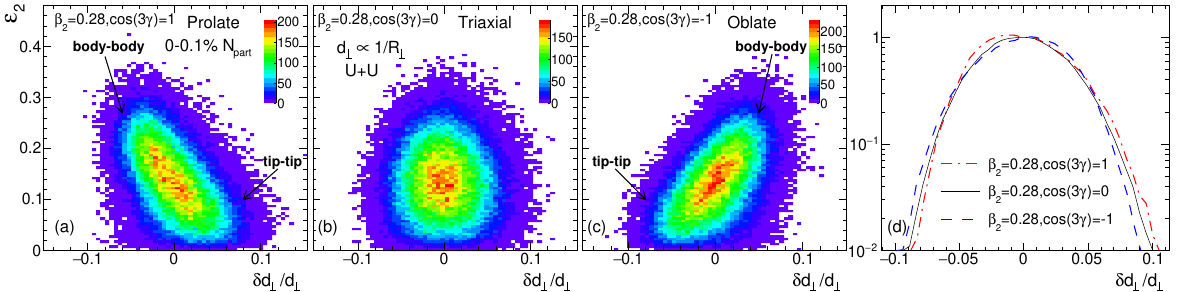}
\end{center}
\caption{\label{fig:2} Correlation between $\varepsilon_2$ and $\delta d_{\perp}/d_{\perp}$ for quadrupole deformation $\beta_2=0.28$ with prolate (left panel), rigid triaxial (second left panel) and oblate (third left panel) shape in the 0--0.1\% most central U+U collisions selected on $\npart$. The right panel show the distributions of $\delta d_{\perp}/d_{\perp}$ in the three cases.}
\end{figure}

To highlight the general feature of the $(\beta_2,\gamma)$ dependence, Fig.~\ref{fig:2} shows the correlations between $\varepsilon_2$ and $\delta d_{\perp}/d_{\perp}=-\delta R_{\perp}/R_{\perp}$~\footnote{In principle full expression should also contain contribution from volume fluctuations, i.e. $\delta d_{\perp}/d_{\perp}=-\delta R_{\perp}/R_{\perp}+\frac{1}{2}\delta \npart /\npart$. However, the second term drops out when one classifies events according to $\npart$.} calculated with nucleon Glauber model in the 0--0.1\% most central U+U collisions selected on $\npart$. They can be contrasted directly with the expectations illustrated by Fig.~\ref{fig:1}.  A clear anticorrelation (positive correlation) between $\varepsilon_2$ and $\delta d_{\perp}/d_{\perp}$ is observed for the prolate (oblate) deformation as expected. The distribution of $\delta d_{\perp}/d_{\perp}$ also indicates clearly a positive (negative) skewness as expected. These distributions are broader than the ideal case in Fig.~\ref{fig:1} due to randomness of $\Omega_1$ relative to $\Omega_2$, surface diffuseness, smearing from nucleon position fluctuations and centrality selection.

The goal of this paper is to explore the $(\beta_2,\gamma)$ dependence of various cumulants in Tabs.~\ref{tab:1} and \ref{tab:2}, and to provide guidance for the experimental measurements. The main finding is that the $\beta_2,\gamma$ dependence for the $n^{\rm{th}}$-order cumulant can be described by a simple equation with the following general form
\begin{align}\label{eq:17}
 a'+(b'+c'\cos(3\gamma)) \beta_2^n\;,
\end{align}
including the variance $\lr{(\delta d_{\perp}/d_{\perp})^2}$ and $\lr{\varepsilon_2^2}$, the skewness $\lr{(\delta d_{\perp}/d_{\perp})^3}$ and $\lr{\varepsilon_2^2\delta d_{\perp}/d_{\perp}}$, and the kurtosis $\lr{(\delta d_{\perp}/d_{\perp})^4}-3\lr{(\delta d_{\perp}/d_{\perp})^2}^2$ and $\lr{\varepsilon_2^4}-2\lr{\varepsilon_2^2}^2$. It is remarkable that most $\gamma$ dependences can be described by a $\cos (3\gamma)$ function, and the higher-order terms allowed by symmetry $\cos (6\gamma)$, $\cos (9\gamma)$ etc are very small. The coefficients $a',b'$ and $c'$ are functions of centrality and collision systems, but are independent of $\beta_2$ and $\gamma$. The coefficient $a'$ represents the values for spherical nuclei, it is usually a strong function of centrality and size of the collision systems. In contrast, the values of $b'$ and $c'$ are similar between nucleon and quark Glauber models and between U+U vs Zr+Zr (i.e. independent of collision systems). They also have rather weak dependence on event centrality. These behavior are the result of geometrical effects: the deformation changes the distribution of nucleons in the entire nucleus, therefore the values of $b'$ and $c'$ in each collision event depend only on the Euler angles of the two nuclei and the impact parameter, and they should be insensitive to the size of the collision system in the Glauber model.

The results are organized as follows. Section~\ref{sec:41} discusses the variance of $d_{\perp}$ in detail, which corresponds to experimentally measured $[\pT]$ variance. Results of higher-order cumulants, skewness and kurtosis of $d_{\perp}$ fluctuations, are presented in Sec.~\ref{sec:42}. Section~\ref{sec:43} considers the mixed cumulant between $d_{\perp}$ and $\varepsilon_2$, which is identified to be the most promising observable to constrain $\gamma$. I then summarize in Sec.~\ref{sec:44} the Glauber results in terms of Eq.~\eqref{eq:17} and discuss the effects of volume fluctuations, and the centrality and system dependences of the results. The results of the AMPT model are included in the Appendix~\ref{sec:app0}.

\subsection{Variance of $d_{\perp}$ fluctuations}\label{sec:41}
In the hydrodynamic picture, the variance of $d_{\perp}$ fluctuation is proportional to the variance of $[\pT]$ fluctuation, $C_{\mathrm{d}}\{2\}=\lr{(\delta d_{\perp}/d_{\perp})^2} \propto \lr{(\delta [\pT]/\lr{\pT})^2}$.
\begin{figure}[h!]
\begin{center}
\includegraphics[width=1\linewidth]{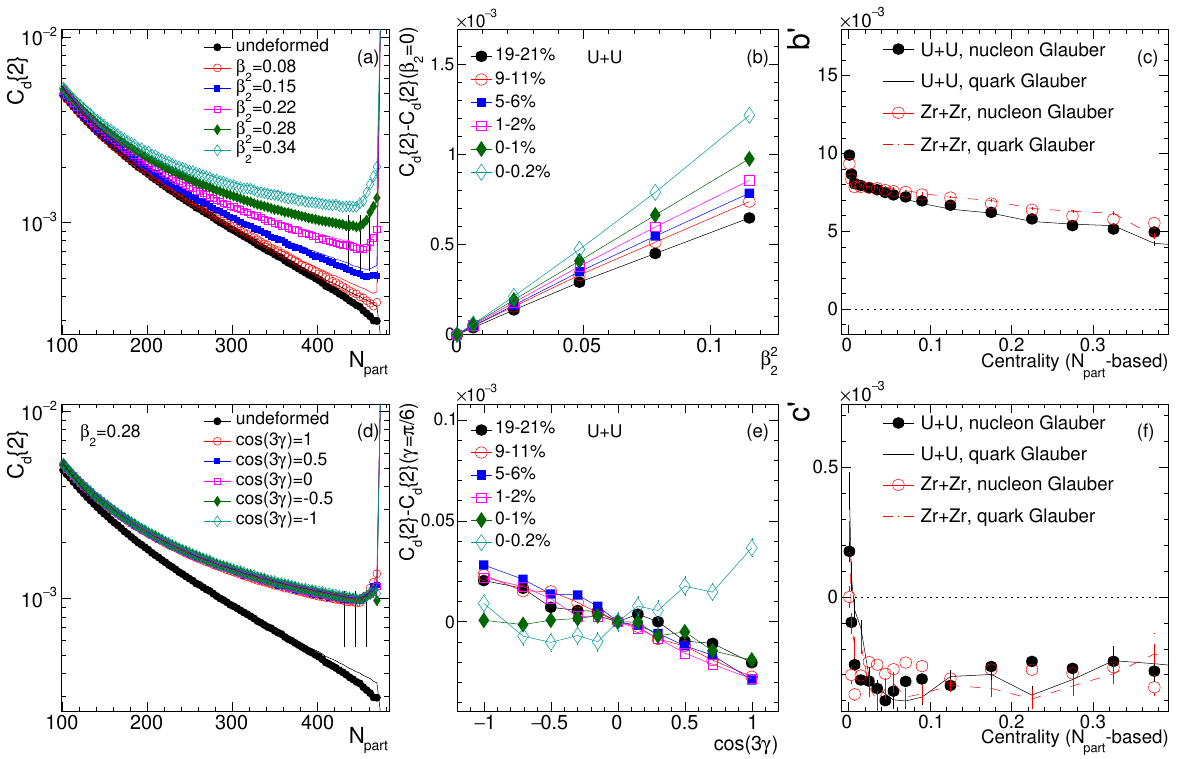}
\end{center}\vspace*{-0.6cm}
\caption{\label{fig:3} $\lr{(\delta d_{\perp}/d_{\perp})^2}$ for several $\beta_2$ values with $\gamma=0$ (top row) and several $\gamma$ values with $\beta_2=0.28$ (bottom row) in U+U collisions. The left column show the $\npart$ dependence where markers and lines represent $d_{\perp}$ obtained with nucleons and quarks, respectively. The middle column shows results in several centrality ranges, which follows a linear function of $\beta_2^2$ (top panel) or $\cos(3\gamma)$ (bottom panel). The right column shows the coefficients $b'$ (top) and $c'$ (bottom) as a function of centrality in U+U (black) and Zr+Zr (red) systems for $d_{\perp}$ calculated from nucleons (markers) or quarks (lines). The three vertical lines in the left column mark the locations of 2\%, 1\% and 0.2\% centrality, respectively.}
\end{figure}
\begin{figure}[h!]
\begin{center}\vspace*{-0.3cm}
\includegraphics[width=1\linewidth]{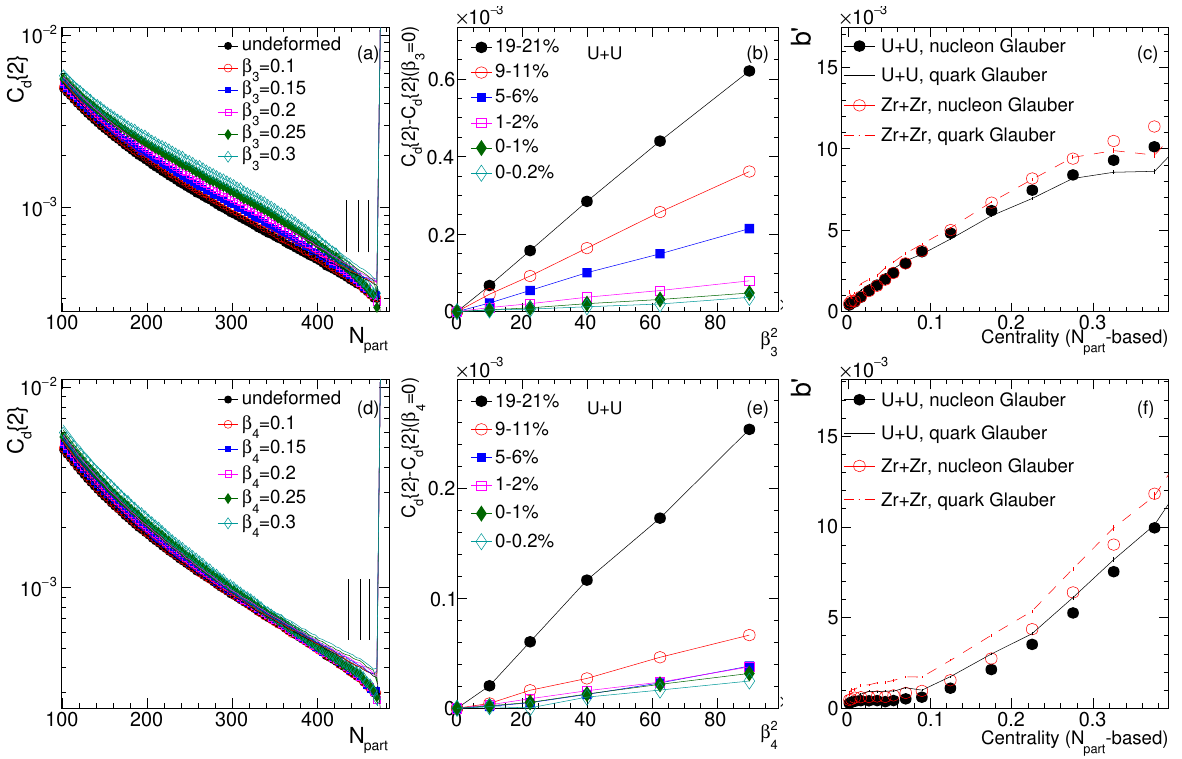}
\end{center}\vspace*{-0.6cm}
\caption{\label{fig:4} $\lr{(\delta d_{\perp}/d_{\perp})^2}$ for several values of $\beta_3$ (top row) and $\beta_4$ (bottom row) as a function of $\npart$ (left column) or $\beta_n^2$ (middle column) in U+U collisions. The latter dependences can be described by a simple $a'+b'\beta_n^2$ function. The right column summarizes the values of $b'$ from the middle column as a function of centrality in U+U (black) and Zr+Zr (red) systems.}
\end{figure}

The left column of Fig.~\ref{fig:3} shows the $\npart$ dependence of $C_{\mathrm{d}}\{2\}$ for various values of $\beta_2$ or $\gamma$ with fixed $\beta_2=0.28$ in U+U collisions, calculated from the participating nucleons. In the absence of deformation, the $C_{\mathrm{d}}\{2\}$ decreases approximately as a power-law function of $\npart$. The presence of large $\beta_2$ increases $C_{\mathrm{d}}\{2\}$ over a very broad centrality range. On the other hand, the triaxiality parameter $\gamma$ only has a small influence, as reflected by the clustering of all different curves in the bottom-left panel. In the same panels, I also show results calculated from quark participants as solid lines, with the same color as those calculated from nucleon participants. Little differences are observed between the two, implying that the influences of deformation are insensitive to nucleon substructures.   
\begin{figure}[h!]
\begin{center}
\includegraphics[width=1\linewidth]{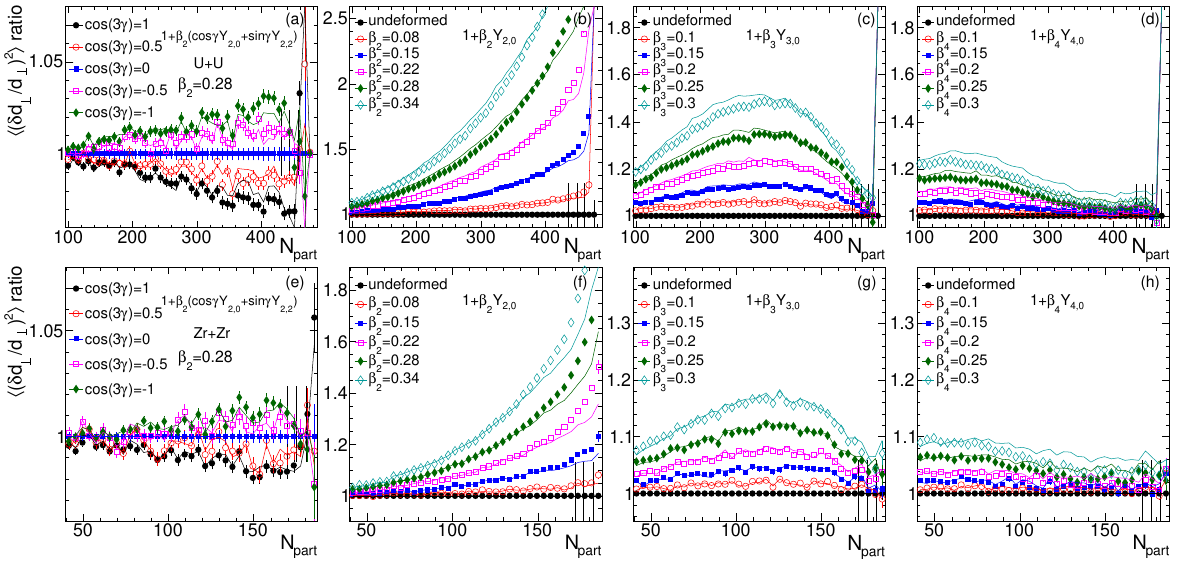}
\end{center}\vspace*{-0.3cm}
\caption{\label{fig:5} Ratios of $\lr{(\delta d_{\perp}/d_{\perp})^2}$ to the default as a function of $\npart$ for several values of $\cos(3\gamma)$ with $\beta_2=0.28$ (left column), several values of $\beta_2$ with $\cos(3\gamma)=1$ (second column), several values of $\beta_3$ (third column) and $\beta_4$ (right column) in the U+U (top row) and the Zr+Zr (bottom row) collisions. The results calculated using nucleons or quarks are shown in markers and lines respectively. The three vertical bars around unity mark the locations of 2\%, 1\% and 0.2\% centrality, respectively.}
\end{figure}

To quantify the $(\beta_2,\gamma)$ dependencies, $C_{\mathrm{d}}\{2\}$ values obtained for fixed $\npart$ are averaged in narrow centrality ranges, which are then plotted as a function of $\beta_2^2$ or $\cos(3\gamma)$ in the middle column of Fig.~\ref{fig:3}. Very good linear trends are observed in most of the cases, confirming Eq.~\eqref{eq:17}~\footnote{In 0--0.2\% centrality one also observes significant $\cos(6\gamma)$ component in Fig.~\ref{fig:3}, but not in quark Glauber model.}. The slopes in the middle-top panel equal to $b'+c'$ (since $\gamma=0$) and the slopes in the middle-bottom panel equal to $c'\beta_2^2$.  The two panels in the right column summarize the centrality dependence of $b'$ and $c'$, respectively. They are shown for $d_{\perp}$ calculated from both nucleons and quarks in U+U and Zr+Zr collisions. It is quite remarkable that the values of $b'$ and $c'$ are insensitive to subnucleon structures and are similar in both collision systems, this is expected since deformation influences the global geometry of the overlap region. The values of $c'$ is about a factor 20--30 smaller than $b'$. A qualitatively similar functional form was also observed between $\lr{\varepsilon_2^2}$ and $(\beta_2,\gamma)$ in a previous study~\cite{Jia:2021tzt}.

Although the axial quadrupole distortion is the nuclear deformation of primary importance, contributions from octupole and hexadecapole components often coexist and can be important in some regions of nuclear chart~\cite{Butler:2016rmu}. Therefore, it is interesting to study how $d_{\perp}$ is affected by $\beta_3$ and $\beta_4$. I have performed such calculations and the results are shown in Fig.~\ref{fig:4} with a similar layout as Fig.~\ref{fig:3}. These higher-order deformations have no influence on the variance of $d_{\perp}$ in the UCC region, but significant enhancement associated with $\beta_3$ is observed in near-central and mid-central collisions, and the $\beta_4$ only has a modest enhancement in the peripheral region. These enhancements can be described by a quadratic function $b'\beta_3^2$ or $b'\beta_4^2$ according to Eq.~\eqref{eq:10}. The coefficients $b'$ are shown in the right panels. 

\begin{figure}[h!]
\begin{center}
\includegraphics[width=0.58\linewidth]{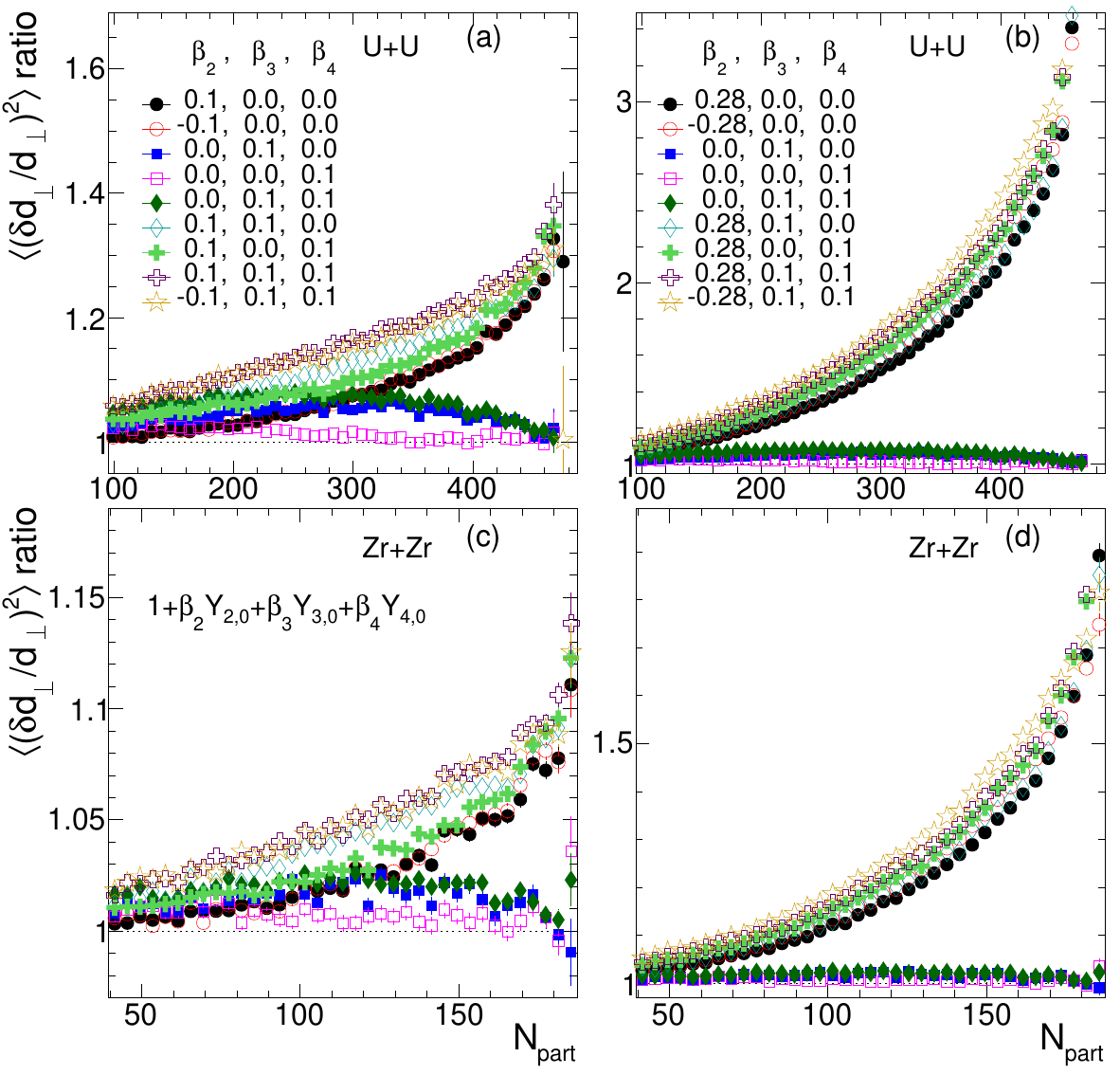}
\end{center}\vspace*{-0.3cm}
\caption{\label{fig:6} Ratio of $\lr{(\delta d_{\perp}/d_{\perp})^2}$ for U+U (top row) and Zr+Zr (bottom row) collisions, relative to spherical case, as a function of $\npart$ for different combinations of $\beta_2$, $\beta_3$ and $\beta_4$ as indicated in the top panels. The left column shows results with small $\beta_2=0.1$, while the right column shows results with large $\beta_2=0.28$. Both $d_{\perp}$ and centrality are based on nucleon participants.}
\end{figure}

To better visualize and quantify the effects of deformation, Fig.~\ref{fig:5} shows the ratios of $C_{\mathrm{d}}\{2\}$ in U+U (top row) or in Zr+Zr (bottom) collisions. The results in the top row are obtained directly from the data from the left columns of Figs.~\ref{fig:3} and \ref{fig:4}. These results can be related to the ratios of $\lr{(\delta [\pT]/[\pT])^2}$ between two systems with similar mass number but different deformation parameters. Most trends are obvious, but the results for different $\gamma$ cases deserve some discussion. The separation between different $\gamma$ cases increases linearly with $\npart$, reaching its maximum around 2\% centrality and then decreasing in the more central region. The maximum relative difference is about 3--4\%, which is about twice of the influence of $\gamma$ for $\varepsilon_2$~\cite{Jia:2021tzt}. As discussed later, such a $\gamma$ dependence may arise from the higher-order expansion of $\delta d_{\perp}/d_{\perp}$ in powers of $\beta_2$, which is particularly important for the kurtosis of the $d_{\perp}$ fluctuations. 

It is also interesting to study how the fluctuations of $d_{\perp}$ depend on the simultaneous presence of quadrupole and higher-order deformations, in particular, whether the contribution from each component to $d_{\perp}$ is independent of each other.  For this exploratory study, only combinations of axial-symmetric components $Y_{n,0}, n=2,3, 4$ are considered. The analysis is carried out for different combinations of $(\beta_2,\beta_3,\beta_4)$ from the values $\beta_2=\pm0.1,0$, $\beta_3=0.1,0$, and $\beta_4=0.1,0$, and results are shown in the left column of Fig.~\ref{fig:6}.  The contributions from different deformation components are not fully independent of each other. In particular, the influence of $\beta_4$ and to some extent also $\beta_3$ is enhanced in the presence of $\beta_2$. This suggests that the mixing between different deformation, i.e. terms such as $\beta_2\beta_4$, $\beta_2\beta_3$ and $\beta_3\beta_4$ in Eq.~\eqref{eq:10} are more important, but these nonlinear effects are always very small in the UCC region. The right column of Fig.~\ref{fig:6} considers a different scenario where the quadrupole component  $\beta_2=0.28$ is much larger than the octupole and hexadecapole. Similar conclusions can be drawn. 

\subsection{Skewness and kurtosis of $d_{\perp}$ fluctuations}\label{sec:42}
\begin{figure}[h!]
\begin{center}
\includegraphics[width=1\linewidth]{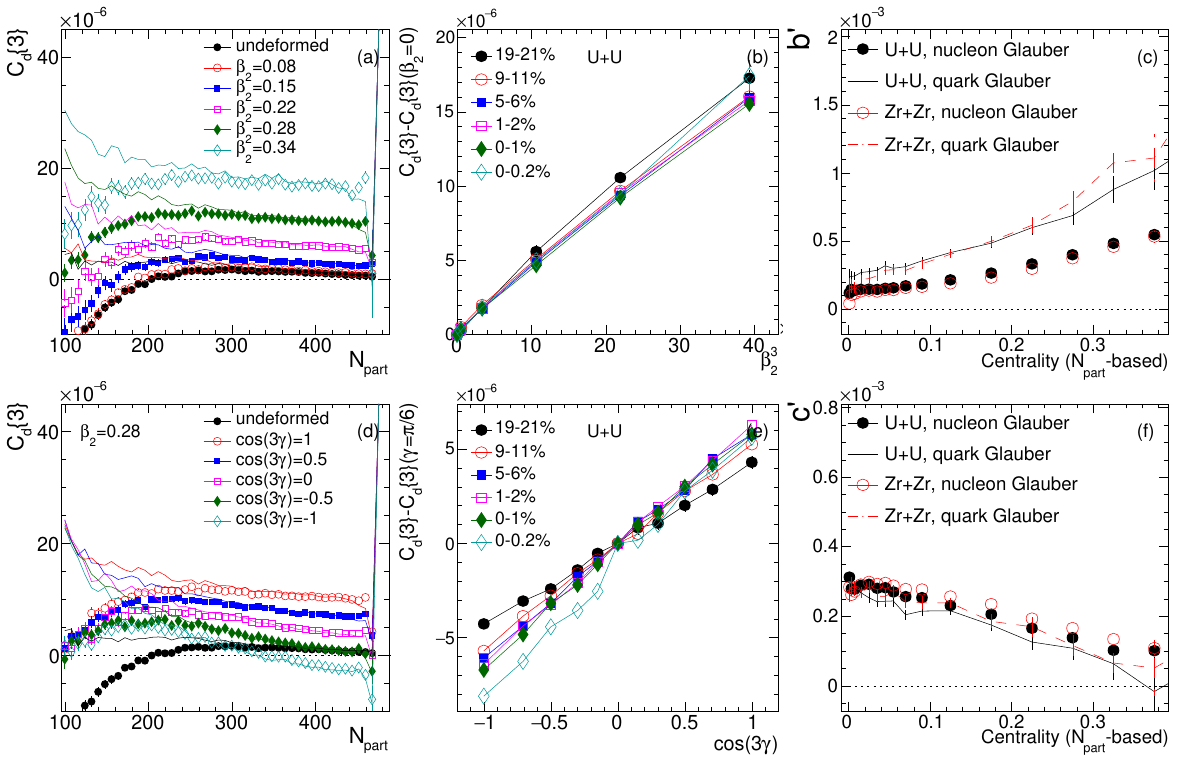}
\end{center}\vspace*{-0.3cm}
\caption{\label{fig:7} The skewness $\lr{(\delta d_{\perp}/d_{\perp})^3}$ for several $\beta_2$ values with $\gamma=0$ (top row) and several $\gamma$ values with $\beta_2=0.28$ (bottom row). The left column shows the $\npart$ dependence where markers and lines correspond $d_{\perp}$ obtained with nucleons and quarks, respectively. The middle column shows the respective results in several centrality ranges based on $\npart$, which can be mostly described by a linear function of $\beta_2^3$ (top panel) or $\cos(3\gamma)$ (bottom panel) via Eq.~\eqref{eq:17}. The right column summarizes extracted coefficients $b'$ (top) and $c'$ (bottom) as a function of centrality in U+U (black) and Zr+Zr (red) systems calculated from nucleons (markers) or quarks (lines).}
\end{figure}

Figure~\ref{fig:7} shows the results of skewness $C_{\mathrm{d}}\{3\}=\lr{(\delta d_{\perp}/d_{\perp})^3}$, which is directly related to the skewness of transverse momentum fluctuations $\lr{(\delta [\pT]/\lr{\pT})^3}$, for different values of $\beta_2$ and $\gamma$ with similar layout as Fig.~\ref{fig:3}.
  The $\npart$ dependence in the left column show a strong sensitivity to the deformation parameter values across a broad centrality range. In particular, the $C_{\mathrm{d}}\{3\}$ in the presence of large $\beta_2$ is nearly constant from the mid-central to central collisions, a salient feature observed in the skewness of $[\pT]$ fluctuations in the U+U data by the STAR collaboration~\cite{jjia}.  The bottom panel also shows that the $C_{\mathrm{d}}\{3\}$ is largest for prolate deformation $\cos(3\gamma)=1$ and smallest for the oblate deformation $\cos(3\gamma)=-1$. In the latter case, $C_{\mathrm{d}}\{3\}$ changes sign to negative in central collisions. The $C_{\mathrm{d}}\{3\}$ values are plotted as a function of $\beta_3^2$ or $\cos(3\gamma)$ in the middle panels. Very good linear dependencies, described by Eq.~\eqref{eq:17}, are observed.

The right panels show the centrality dependence of the coefficients $b'$ and $c'$ for various cases. The results are similar between U+U and Zr+Zr collisions, but the values of $b'$ obtained from quark Glauber model are systematically larger, especially towards more peripheral collisions. The values of $c'$ are larger than $b'$ in the 0\%--10\% most central collisions, and are smaller than $b'$ in the mid-central and peripheral collisions. This should be contrasted to the expectation of liquid-drop model, which predicts $b'=0$ in the UCC region. The strong sensitivity to $\gamma$ suggests that the skewness of the $[\pT]$ fluctuation is an excellent probe of nuclear triaxiality. For smaller Zr+Zr collision system, one does not observe a sign change from prolate deformation to oblate deformation even with $\beta_2=0.28$ (see Fig.~\ref{fig:app2} in Supplemental Material)

\begin{figure}[h!]
\begin{center}
\includegraphics[width=1\linewidth]{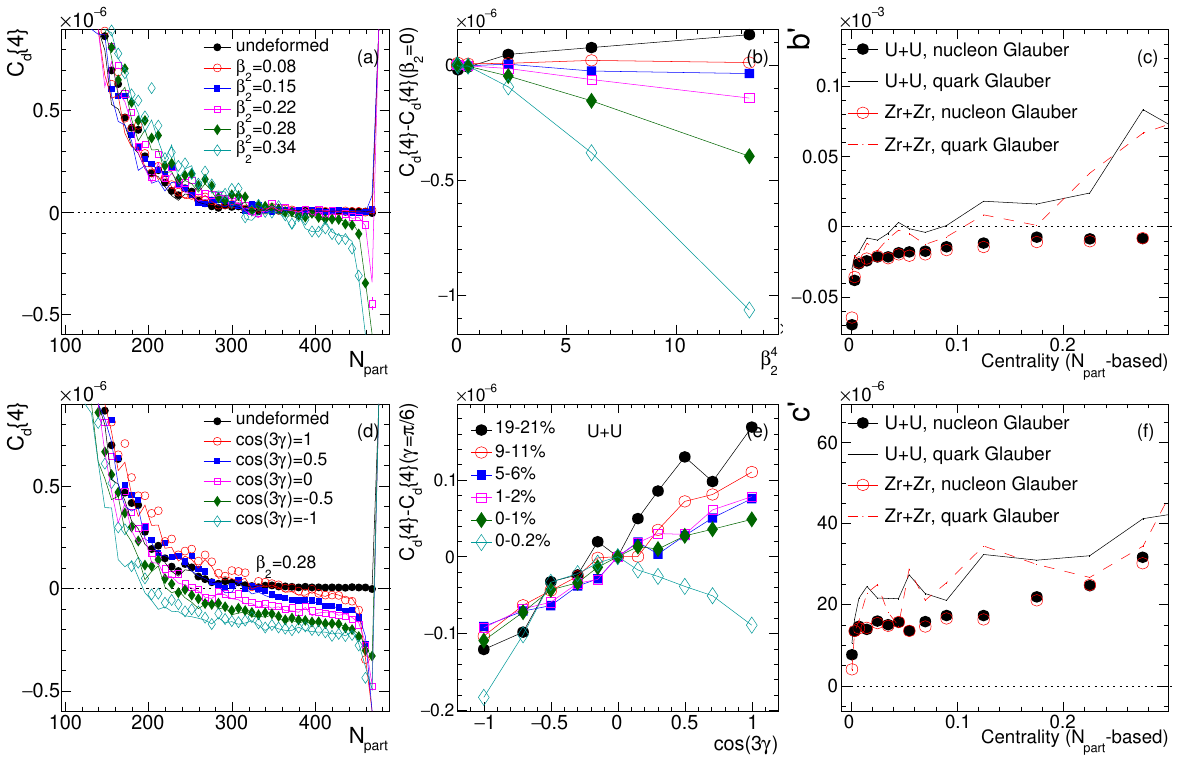}
\end{center}\vspace*{-0.3cm}
\caption{\label{fig:8} The kurtosis of $p(d_{\perp})$ for several $\beta_2$ values with $\gamma=0$ (top row) and several $\gamma$ values with $\beta_2=0.28$ (bottom row). The left column shows the $\npart$ dependence where markers and lines correspond $d_{\perp}$ obtained with nucleons and quarks, respectively. The middle column shows the respective results in several centrality ranges based on $\npart$, which can be mostly described by a linear function of $\beta_2^4$ (top panel) or $\cos(3\gamma)$ (bottom panel). The right column shows centrality dependence of extracted $b'$ (top) and $c'$ (bottom) via Eq.~\eqref{eq:17} in U+U (black) and Zr+Zr (red) systems for $d_{\perp}$ calculated from nucleons (markers) or quarks (lines).}
\end{figure}

Results for kurtosis $C_{\mathrm{d}}\{4\}=\lr{(\delta d_{\perp}/d_{\perp})^4}-3\lr{(\delta d_{\perp}/d_{\perp})^2}^2$ are shown Fig.~\ref{fig:8}, which can be used to provide guidance on the behavior of kurtosis of transverse momentum fluctuations $\lr{(\delta [\pT]/\lr{\pT})^4}-3\lr{(\delta [\pT]/\lr{\pT})^2}^2$. For large prolate deformation (top row), $C_{\mathrm{d}}\{4\}$ changes sign in the UCC region. It also shows a strong dependence on $\gamma$ (bottom row), i.e. $C_{\mathrm{d}}\{4\}$ becomes more negative when nuclear shape change from probate to oblate. These dependencies again can be parametrized according to Eq.~\eqref{eq:17}. The centrality dependence of the extracted coefficients $b'$ and $c'$ are shown in the right panels. Besides the similarity between U+U and Zr+Zr, one finds $b'\approx -c'$ in the case of nucleon Glauber model, but $|b'|\gg|c'|$ in the quark Glauber model. The origin for this is related to a small $\cos(3\gamma)$ dependence in the $C_{\mathrm{d}}\{2\}$, which will be discussed later.

\begin{figure}[h!]
\begin{center}
\includegraphics[width=0.49\linewidth]{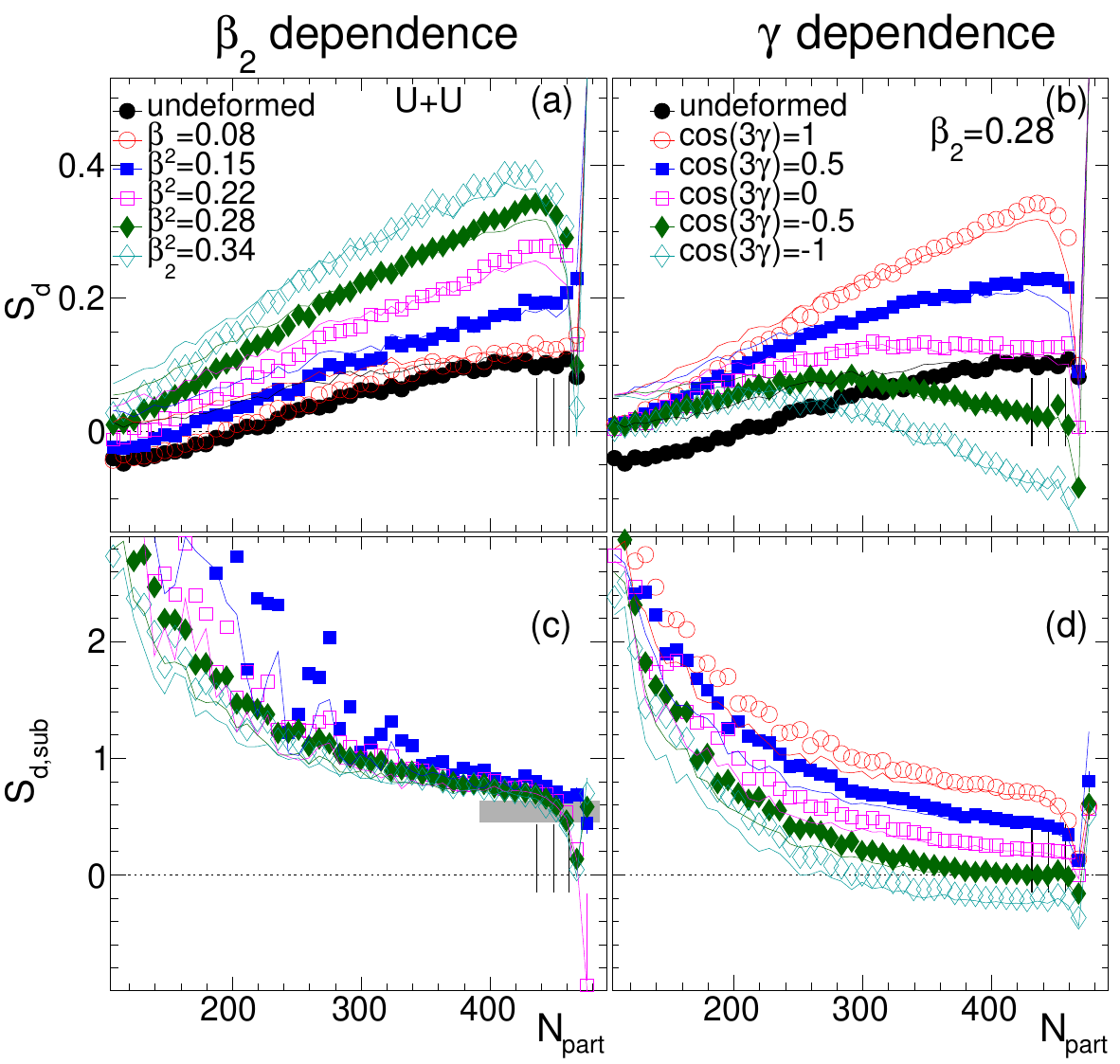}\includegraphics[width=0.49\linewidth]{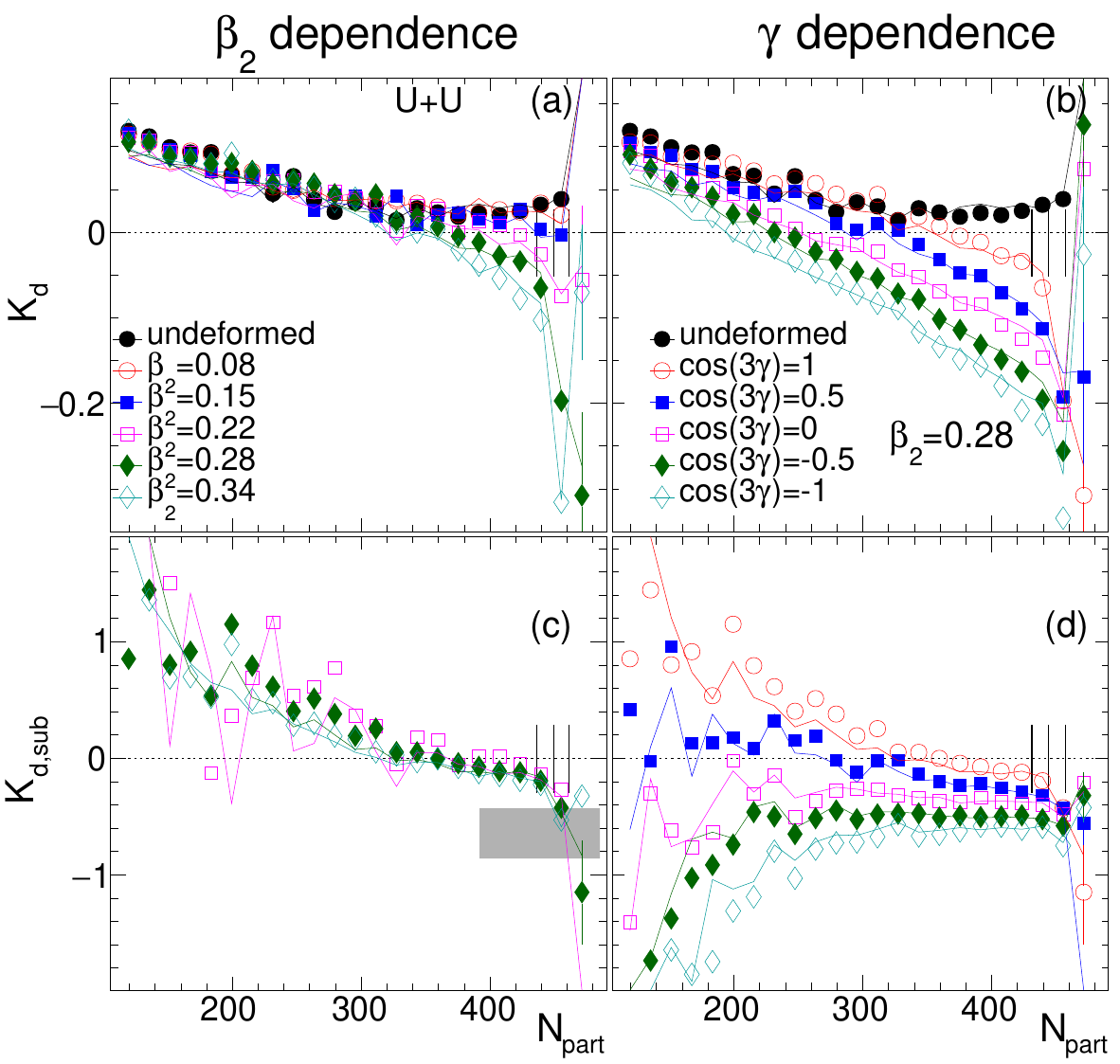}
\end{center}\vspace*{-0.3cm}
\caption{\label{fig:9} Left part: $\npart$ dependence of normalized skewness $S_{\mathrm{d}}=C_{\mathrm{d}}\{3\}/C_{\mathrm{d}}\{2\}^{3/2}$ (top) and modified version $S_{\mathrm{d,sub}}$ (bottom) for several $\beta_2$ values with $\gamma=0$ (left) and for several $\gamma$ values with $\beta_2=0.28$ (right) in U+U collisions. Right part: results for normalized kurtosis $K_{\mathrm{d}}=C_{\mathrm{d}}\{4\}/C_{\mathrm{d}}\{2\}^{2}$ and $K_{\mathrm{d,sub}}$ with the same layout. The shaded bands indicate the predicted range from Tabs.~\ref{tab:1} and \ref{tab:2}.}
\end{figure}

The behavior of the high-order cumulants are often analyzed in terms of cumulant ratios. In an independent source picture and without deformation, the cumulants of intensive quantities scales approximately as $C_{\mathrm{d}}\{k\}\sim 1/\npart^{k-1}$. The normalized skewness $S_{\mathrm{d}}$ and normalized kurtosis $K_{\mathrm{d}}$ in Eq.~\eqref{eq:13} are expected to scale naively as $S_{\mathrm{d}} \sim 1/\sqrt{\npart}$ and $K_{\mathrm{d}} \sim 1/\npart$, respectively. The results of Glauber model using $\npart$-based event averaging in Fig.~\ref{fig:7} show clear deviation from this scaling expectation, although results obtained using $\nqp$-based event averaging are closer to this scaling.  The presence of nuclear deformation is expected to cause further deviation from this baseline. The top row of Fig.~\ref{fig:9} shows the $S_{\mathrm{d}}$ (left two panels) and $K_{\mathrm{d}}$ (right two panels) as a function of $\npart$ for various $\beta_2$ and $\gamma$ values. In the presence of large $\beta_2$, the values of $S_{\mathrm{d}}$ are greatly enhanced, while the values of $K_{\mathrm{d}}$ decrease more strongly and even change sign in the UCC region. As one varies $\gamma$ from prolate to oblate with fixed $\beta_2=0.28$, the behavior of $S_{\mathrm{d}}$ changes from an increase with $\npart$ to a decrease with $\npart$, while $K_{\mathrm{d}}$ deceases nearly linearly with $\npart$ with an increasingly larger slope. The results of $K_{\mathrm{d}}$ suggest a fairly sizable $\cos(3\gamma)$ component on the order of 0.1--0.2. As mentioned earlier, the origin is related to the residual $\cos(3\gamma)$ dependence in the $C_{\mathrm{d}}\{2\}$ in Fig.~\ref{fig:5}. This small $\gamma$ dependence at the level of $\Delta C_{\mathrm{d}}\{2\}/C_{\mathrm{d}}\{2\}\approx\pm0.03$ is found to contribute to the kurtosis approximately as $\Delta K_{\mathrm{d}} =\frac{\Delta C_{\mathrm{d}}\{4\}}{C_{\mathrm{d}}\{2\}^2}-6\frac{\Delta C_{\mathrm{d}}\{2\}}{C_{\mathrm{d}}\{2\}} \approx -4\frac{\Delta C_{\mathrm{d}}\{2\}}{C_{\mathrm{d}}\{2\}} \approx \mp0.12$.

The normalized skewness $S_{\mathrm{d}}$ and kurtosis $K_{\mathrm{d}}$, while easier to construct experimentally, mix up the contributions from nucleon fluctuations and nuclear deformation, which preclude a direct and intuitive interpretation of the results. Therefore, I propose a modified form of the normalized cumulants,
\begin{align}\label{eq:18}
S_{\mathrm{d,sub}} \equiv \frac{C_{\mathrm{d}}\{3\}-C_{\mathrm{d}}\{3\}_{|\beta_2=0}}{(C_{\mathrm{d}}\{2\}-C_{\mathrm{d}}\{2\}_{|\beta_2=0})^{3/2}} \equiv S_{\mathrm{d}}(\beta_2=\infty) \;,\;K_{d.sub}  \equiv \frac{C_{\mathrm{d}}\{4\}-C_{\mathrm{d}}\{4\}_{|\beta_2=0}}{(C_{\mathrm{d}}\{2\}-C_{\mathrm{d}}\{2\}_{|\beta_2=0})^{2}} \equiv K_{\mathrm{d}}(\beta_2=\infty)
\end{align}
With this definition, the baseline contributions are subtracted in the numerator and denominator and the $\beta_2$ dependence is expected to cancel. The final results contain only the $\cos(3\gamma)$ dependence and can be compared directly with the normalized quantities in Tables~\ref{tab:1} and \ref{tab:2}. Another important point is that the values of the normalized cumulant are expected to lie in between two limits
\begin{align}\label{eq:19}
S_{\mathrm{d}}(\beta_2=0)<S_{\mathrm{d}}(\beta_2)<S_{\mathrm{d,sub}}\;,\;\;K_{\mathrm{d,sub}}<K_{\mathrm{d}}(\beta_2)<K_{\mathrm{d}}(\beta_2=0)\;.
\end{align}

The bottom panels of Fig.~\ref{fig:9} show the results for these modified quantities. Results for different $\beta_2$ values, as shown by the first panel for $S_{\mathrm{d,sub}}$ and the third panel for $K_{\mathrm{d,sub}}$, nearly collapse on a common curve, confirming the earlier statement that these observables are a great way to isolate the coefficient $b'$ and $c'$ in Eq~\eqref{eq:17}. The same panels also show the range of the predicted values from Tabs.~\ref{tab:1} and \ref{tab:2} by the shaded gray boxes. Remarkably, the values predicted from the full Monte Carlo Glauber model falls within the ranges from the simple analytical estimates. These results suggest an approximate parametrization $S_{\mathrm{d,sub}}=b_0+c_0\cos3\gamma$, with coefficient $c_0$ nearly independent of centrality and coefficient $b_0$ increasing from central to peripheral collisions.

Even though $S_{\mathrm{d,sub}}$ and $K_{\mathrm{d.sub}}$ can not be directly measured, they can be estimated by comparing results from collisions of two species $A$ and $B$ with similar mass numbers. Taking the skewness for example, one could construct the following ratio using Eq.~\eqref{eq:17},
\begin{align}\label{eq:20}
S_{\mathrm{d,AB}}=\frac{C_{\mathrm{d}}\{3\}_{\mathrm{A}}-C_{\mathrm{d}}\{3\}_{\mathrm{B}}}{(C_{\mathrm{d}}\{2\}_{\mathrm{A}}-C_{\mathrm{d}}\{2\}_{\mathrm{B}})^{3/2}}  \approx S_{\mathrm{d,sub,A}}\left(1+\frac{3}{2}x^2-\frac{b'+c'\cos(3\gamma_{\mathrm{B}})}{b'+c'\cos(3\gamma_{\mathrm{A}})}x^3+\frac{15}{8}x^4\right)\;,
\end{align}
where $x=\beta_{\mathrm{2B}}/\beta_{\mathrm{2A}}\ll 1$ is assumed and I have ignored the negligible $\cos(3\gamma)$ term in $C_{\mathrm{d}}\{2\}$. The $b'$ and $c'$ refers those of $C_{\mathrm{d}}\{3\}$, which are expected to be the same for the two species. The ideal case for Eq.~\eqref{eq:20} is between a pair of isobaric system with different amount of deformations such as $^{96}$Zr+$^{96}$Zr and $^{96}$Ru+$^{96}$Ru collisions~\cite{STAR:2021mii}.

\subsection{Correlation between eccentricity and $d_{\perp}$}\label{sec:43}
Let us turn our attention to the skewness $\lr{\varepsilon_2^2(\delta d_{\perp}/d_{\perp})}$ and the related final-state observable $\lr{v_2^2(\delta [\pT]/[\pT])}$. This observable has been studied both experimentally~\cite{jjia,ATLAS-CONF-2021-001} and in models~\cite{Giacalone:2019pca,Jia:2021wbq}, and, as discussed below, it has great potential in constraining the triaxiality of the colliding nuclei.

\begin{figure}[h!]
\begin{center}
\includegraphics[width=1\linewidth]{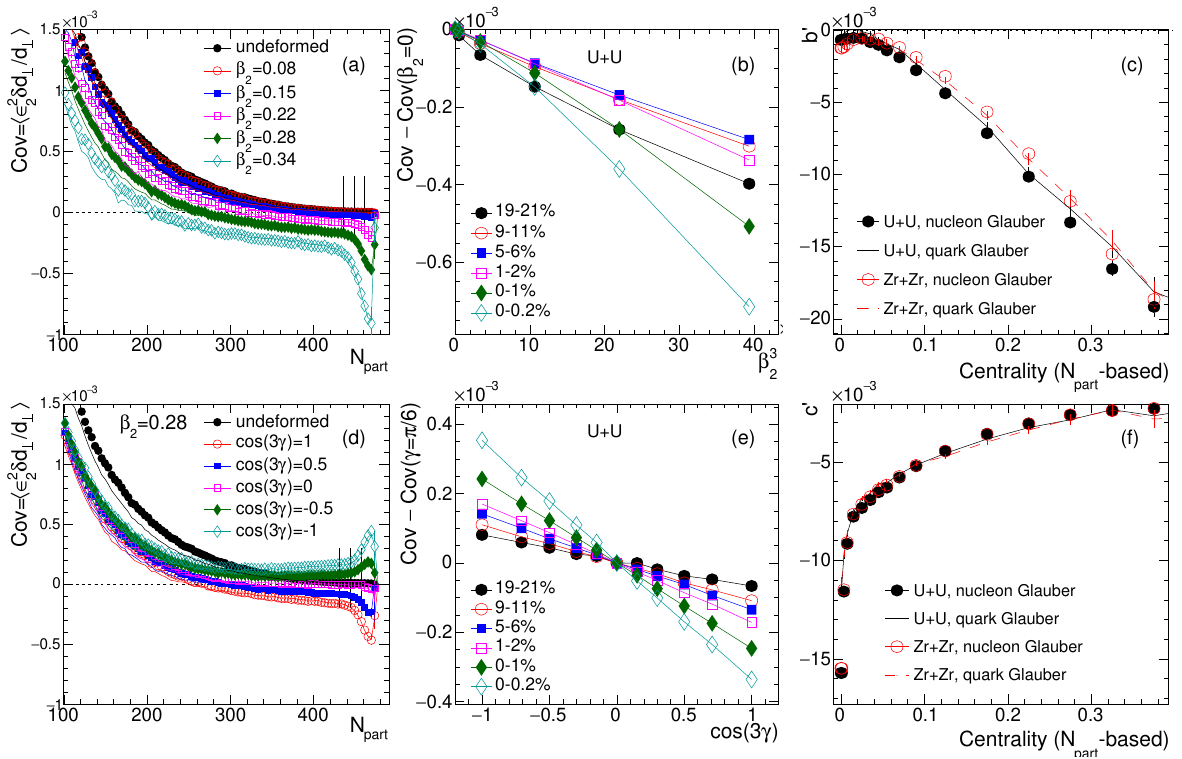}
\end{center}\vspace*{-0.3cm}
\caption{\label{fig:10} The $\lr{\varepsilon_2^2\delta d_{\perp}/d_{\perp}}$ for several $\beta_2$ values with $\gamma=0$ (top row) and several $\gamma$ values with $\beta_2=0.28$ (bottom row). The left column shows the $\npart$ dependence. The middle column shows the respective results in several centrality ranges based on $\npart$. The right column summarizes the centrality dependence of $b'$ (top) and $c'$ (bottom) obtained via Eq.~\eqref{eq:17} in U+U (black) and Zr+Zr (red) collisions for $\varepsilon_2$ and $d_{\perp}$ calculated from nucleons (markers) or quarks (lines).}
\end{figure}

Figure~\ref{fig:10} shows the results of $\lr{\varepsilon_2^2(\delta d_{\perp}/d_{\perp})}$ for different values of $\beta_2$ and $\gamma$ with the usual layout. The $\npart$ dependences show a clear hierarchy between different $\beta_2$ and/or $\gamma$ values, and the sensitivity to these parameters are clearly visible across a broad centrality range. In the absence of deformation, $\lr{\varepsilon_2^2(\delta d_{\perp}/d_{\perp})}$ decreases gradually from peripheral to more central collisions but remains positive. For prolate deformation, as $\beta_2$ is increased, $\lr{\varepsilon_2^2(\delta d_{\perp}/d_{\perp})}$ decreases over the entire centrality range, and becomes negative in the central region. However, for large oblate deformation, $\lr{\varepsilon_2^2(\delta d_{\perp}/d_{\perp})}$ increases in the central region. This behavior is fully consistent with the expectation from Fig.~\ref{fig:1}. 

The middle column shows the values of $\lr{\varepsilon_2^2(\delta d_{\perp}/d_{\perp})}$ as a function of either $\beta_2^3$ or $\cos(3\gamma)$ in several narrow centrality ranges. A linear dependence is observed, consistent with the now familiar parametetrization Eq.~\eqref{eq:17}. The right panels show the centrality dependencies of $b'$ and $c'$ for various cases. The results are similar between U+U and Zr+Zr collisions and between nucleon Glauber and quark Glauber models. Both $b'$ and $c'$ are negative over the full centrality range. But the magnitude of $c'$ is much larger than $b'$ in the 0\%--10\% central collisions, and is smaller than $b'$ in the mid-central and peripheral collisions. The sensitivity of $\lr{\varepsilon_2^2(\delta d_{\perp}/d_{\perp})}$ to $\gamma$ is stronger than $\lr{(\delta d_{\perp}/d_{\perp})^3}$, even though they are clearly complementary~\footnote{Given the importance of this observable, I also investigated the influence of $\beta_3$ and $\beta_4$ (see Fig.~\ref{fig:app6} in Appendix~\ref{sec:app2}). The influence is negligible in the UCC region. But one finds that $\beta_3$ enhances the value of $\lr{\varepsilon_2^2(\delta d_{\perp}/d_{\perp})}$ in central collisions. In the peripheral region, both $\beta_3$ and $\beta_4$ reduce the signal, the relative change is less than 30\% as long as $\beta_3,\beta_4<0.2$.}

\begin{figure}[h!]
\begin{center}
\includegraphics[width=1\linewidth]{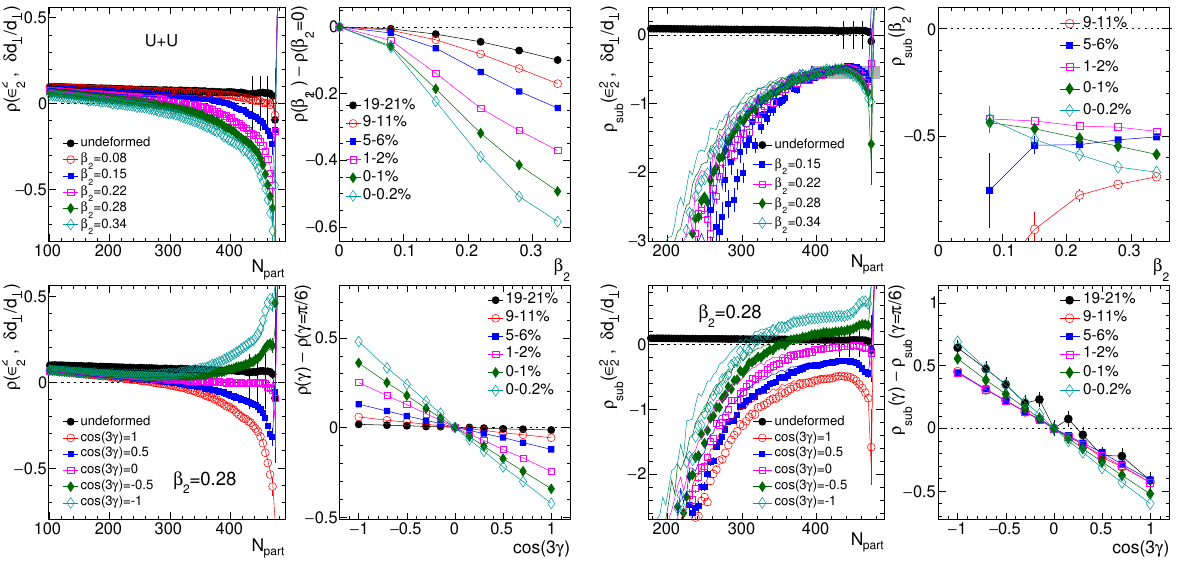}
\end{center}\vspace*{-0.3cm}
\caption{\label{fig:11}  Left part: The left column shows $\rho(\varepsilon_2^2,\delta d_{\perp}/d_{\perp})$ for several $\beta_2$ values with $\gamma=0$ (top row) and and several $\gamma$ values with $\beta_2=0.28$ (bottom row). The left column shows the $\npart$ dependence. The right column shows the $\beta_2$ (top panel) and $\cos(3\gamma)$ (bottom panel) dependencies. Right part: similar plots for $\rho_{\mathrm{sub}}$, and the shaded band in the top-left panel indicate the predicted range from Tabs.~\ref{tab:1} and \ref{tab:2}.}
\end{figure}

The behavior of $\lr{\varepsilon_2^2(\delta d_{\perp}/d_{\perp})}$ can be analyzed using the normalized quantity, $\rho_{\mathrm{orig}}(\varepsilon_2^2,\delta d_{\perp}/d_{\perp})$ and $\rho(\varepsilon_2^2,\delta d_{\perp}/d_{\perp})$ defined in Eq.~\eqref{eq:14a}. They are directly related to the analog experimentally-accessible observable $\rho_{\mathrm{orig}}(v_2^2,\delta [\pT]/[\pT])$~\cite{Bozek:2016yoj} and $\rho(v_2^2,\delta [\pT]/[\pT])$. The results of $\rho(\varepsilon_2^2,\delta d_{\perp}/d_{\perp})$ are shown in the left part of Fig.~\ref{fig:11}. The second column shows an approximately linear function of $\beta_2$ for moderate value of $\beta_2$, but nonlinear behavior shows up at small and larger $\beta_2$. The reason for this complex $\beta_2$ dependence can be attributed to the $a'$ terms in the numerator and the denominator. Following the example for the $S_{\mathrm{d,sub}}$, I define a modified correlator by subtracting out the baseline effects, 
\begin{align}\label{eq:21}
\rho_{\mathrm{sub}}(\varepsilon_2^2,\frac{\delta d_{\perp}}{d_{\perp}}) =\frac{\lr{\varepsilon_2^2 \frac{\delta d_{\perp}}{d_{\perp}}}-\lr{\varepsilon_2^2 \frac{\delta d_{\perp}}{d_{\perp}}}_{|\beta_2=0}}{\left(\lr{\varepsilon_2^2}-\lr{\varepsilon_2^2}_{|\beta_2=0}\right)\sqrt{\lr{(\frac{\delta d_{\perp}}{d_{\perp}})^2}-\lr{(\frac{\delta d_{\perp}}{d_{\perp}})^2}_{|\beta_2=0}}}\equiv\rho(\varepsilon_2^2,\frac{\delta d_{\perp}}{d_{\perp}})_{|\beta_2=\infty}\;, 
\end{align}

Just like the case for skewness of the $d_{\perp}$ fluctuations, the $\beta_2$ dependence completely cancels, and $\rho_{\mathrm{sub}}$ contains only the $\cos(3\gamma)$ dependence. Therefore it can be compared directly to the values in Tabs.~\ref{tab:1} and \ref{tab:2}. The $\rho$ in general is expected to be in between the value without deformation $\rho_{_{|\beta_2=0}}$ and $\rho_{\mathrm{sub}}$.

The right part of Fig.~\ref{fig:11} shows the results for $\rho_{\mathrm{sub}}$. Results for different $\beta_2$ values nearly collapse on a common curve, confirming the earlier statement that these modified quantities are a great way to separate the coefficient $b'$ and $c'$. The same panels also show the range of the predicted values from Tables~\ref{tab:1} and \ref{tab:2}. Remarkably, the values from the full Monte Carlo Glauber model agree well with my analytical estimates. The results suggest $\rho_{\mathrm{sub}}=b_0+c_0\cos3\gamma$, with $c_0$ nearly independent of centrality, while $b_0$ is roughly constant in 0\%--5\% centrality and but decreases beyond that.

Repeating the same argument made for $S_{\mathrm{d,sub}}$, the value of $\rho_{\mathrm{sub}}$ can be estimated by comparing collisions of two species $A$ and $B$ with similar mass number, therefore canceling the baseline effects. The result is,
\begin{align}\label{eq:22}
\rho_{\mathrm{AB}} &=\frac{\lr{\varepsilon_2^2 \frac{\delta d_{\perp}}{d_{\perp}}}_{\mathrm{A}}-\lr{\varepsilon_2^2 \frac{\delta d_{\perp}}{d_{\perp}}}_{\mathrm{B}}}{(\lr{\varepsilon_2^2}_{\mathrm{A}}-\lr{\varepsilon_2^2}_{\mathrm{B}})\sqrt{\lr{(\frac{\delta d_{\perp}}{d_{\perp}})^2}_{\mathrm{A}}-\lr{(\frac{\delta d_{\perp}}{d_{\perp}})^2}_{\mathrm{B}}}}  \approx \rho_{\mathrm{sub,A}}(\varepsilon_2^2,\frac{\delta d_{\perp}}{d_{\perp}}) (1+\frac{3}{2}x^2-\frac{b'+c'\cos(3\gamma_{\mathrm{B}})}{b'+c'\cos(3\gamma_{\mathrm{A}})}x^3+\frac{15}{8}x^4)\;,
\end{align}
where I assume $x=\beta_{\mathrm{2B}}/\beta_{\mathrm{2A}}\ll 1$ and I have ignored the small $\cos(3\gamma)$ terms in $C_{\mathrm{d}}\{2\}$ and $\lr{\varepsilon_2^2}$. The $b'$ and $c'$ are the coefficients for $\lr{\varepsilon_2^2 \frac{\delta d_{\perp}}{d_{\perp}}}$, which are also expected to be the same for the two nuclei. This approximation is accurate within 5\% for $x<0.5$, and the contribution from $x^3$ and $x^4$ terms is less than 5\% for $x<0.3$ (the same also applies for Eq.~\eqref{eq:20}). They can best done for a pair of isobaric system such as Zr+Zr and Ru+Ru collisions, but could also be used for comparison between Au+Au and U+U systems~\footnote{A small correction is required to precisely cancel the $a'$ term~\cite{Giacalone:2021udy}. This can be achieved by focusing on central events with similar multiplicity, where the values of $a'$ are smallest and similar between the two systems.}.

\begin{figure}[h!]
\begin{center}
\includegraphics[width=1\linewidth]{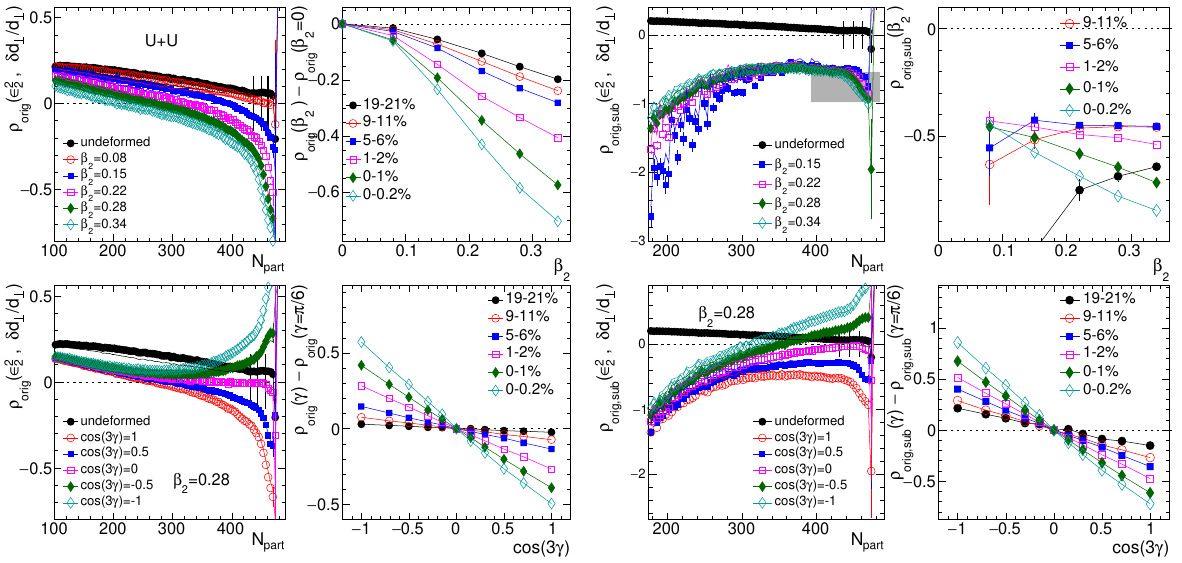}
\end{center}\vspace*{-0.3cm}
\caption{\label{fig:12}  Same as Fig.~\ref{fig:12} but calculated for Pearson correlation coefficients $\rho_{\mathrm{orig}}(\varepsilon_2^2,\delta d_{\perp}/d_{\perp})$ defined in Eq.~\eqref{eq:14a}.}
\end{figure}

Although I do not prefer the standard normalization $\rho_{\mathrm{orig}}(\varepsilon_2^2,\delta d_{\perp}/d_{\perp})$ for deformation studies, I nevertheless carried out the same calculation since it is widely used before. Here the correlator with the baseline effects subtracted is defined as
\begin{align}
\rho_{\mathrm{orig,sub}}(\varepsilon_2^2,\frac{\delta d_{\perp}}{d_{\perp}}) =\frac{\lr{\varepsilon_2^2 \frac{\delta d_{\perp}}{d_{\perp}}}-\lr{\varepsilon_2^2 \frac{\delta d_{\perp}}{d_{\perp}}}_{\beta_2=0}}{\sqrt{(\lr{\left(\delta \varepsilon_2^2\right)^2}-\lr{\left(\delta \varepsilon_2^2\right)^2}_{\beta_2=0})(\lr{(\frac{\delta d_{\perp}}{d_{\perp}})^2}-\lr{(\frac{\delta d_{\perp}}{d_{\perp}})^2}_{\beta_2=0})}}\equiv\rho_{\mathrm{orig}}(\varepsilon_2^2,\frac{\delta d_{\perp}}{d_{\perp}})_{\beta_2=\infty}\;, 
\end{align}
I shall present the final results in Fig.~\ref{fig:12} without detailed discussion. The values and trends in the UCC region are quantitatively similar to $\rho_{\mathrm{sub}}$. This is expected since in central collisions, $c_{2,\varepsilon}\{4\}$ approaches zero and $\lr{\left(\delta \varepsilon_2^2\right)^2}\approx \lr{\varepsilon_2^2}^2$, therefore, $\rho_{\mathrm{orig,sub}}\approx\rho_{\mathrm{sub}}$. In the more peripheral region, the two correlators are quantitatively different. The $\rho_{\mathrm{orig,sub}}$ is relatively flat towards mid-central collisions for prolate deformation with different $\beta_2$, however, its $\gamma$ dependence is much weaker than that for $\rho_{\mathrm{sub}}$.
\begin{figure}[h!]
\begin{center}
\includegraphics[width=0.95\linewidth]{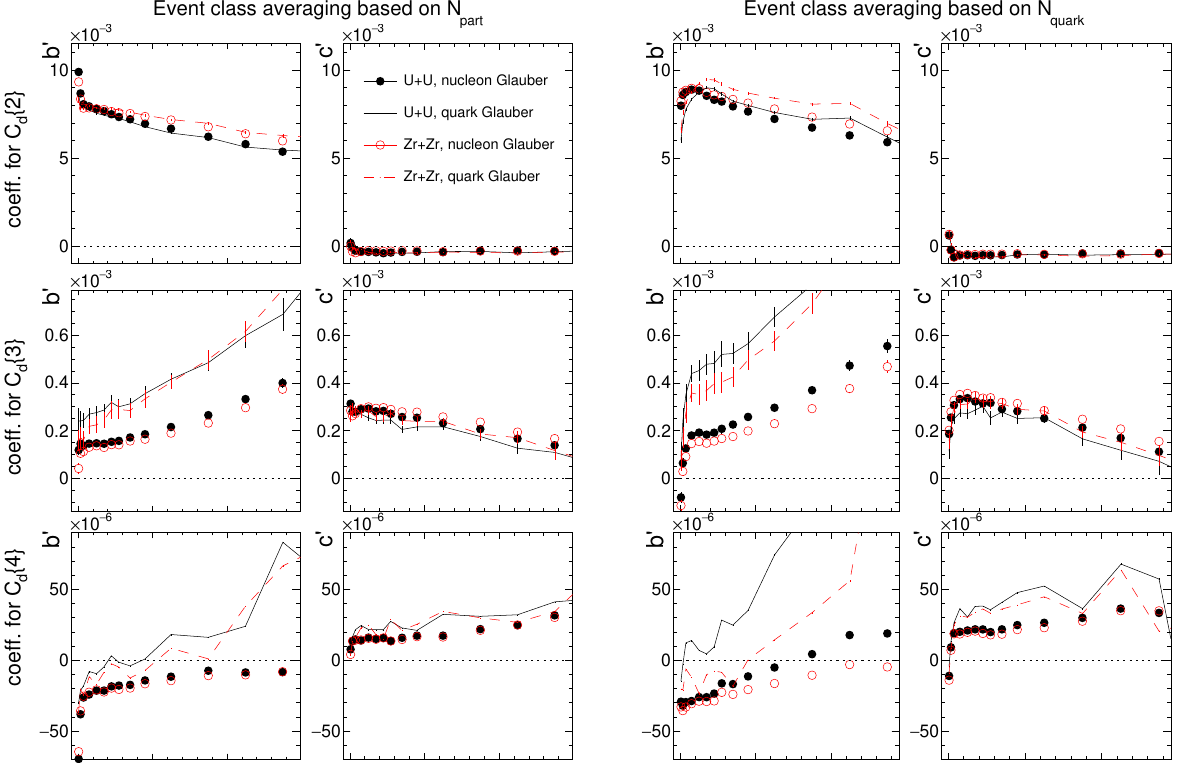}
\includegraphics[width=0.95\linewidth]{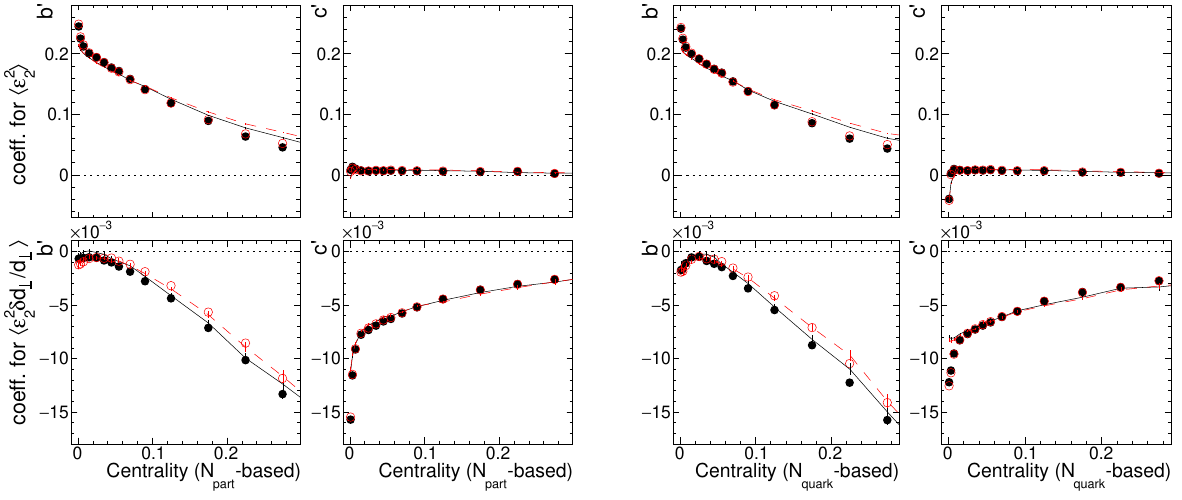}
\end{center}
\caption{\label{fig:13} The centrality dependence of the coefficients $b'$ and $c'$ from Eq.~\eqref{eq:17} for $C_{\mathrm{d}}\{2\}$,$C_{\mathrm{d}}\{3\},C_{\mathrm{d}}\{4\}$, $\lr{\varepsilon_2^2}$ and $\lr{\varepsilon_2^2\delta d_{\perp}/d_{\perp}}$ from the top row to the bottom row. In each row, the values obtained via event averaging based on $\npart$ (left two columns) and $\nqp$ (right two columns) are shown. In each panel, the results are compared between U+U and Zr+Zr, and between values calculated from nucleons (symbols) and quarks (lines). The data points for the following centrality ranges are plotted from left to right: 0\%--0.2\%, 0.2\%--0.5\%, 0.5\%--1\%, 1\%--2\%,..., 5\%--6\%, 6\%--8\%, 8\%--10\%, 10\%--15\%,..., and 25\%--30\%.}
\end{figure}

\subsection{Effects of volume fluctuations and dependence on centrality and system size}\label{sec:44}
Although $d_{\perp}$ and  $\varepsilon_2$ in each event are calculated using either nucleons or quarks, the cumulants of these quantities so far are obtained via an event averaging procedure based on $\npart$. As mentioned before, the averaging could also be performed over event ensembles classified via $\nqp$. Figure~\ref{fig:13} summarizes the coefficients $b'$ and $c'$ as a function of centrality for the five quantities $C_{\mathrm{d}}\{2\},C_{\mathrm{d}}\{3\},C_{\mathrm{d}}\{4\}$, $\lr{\varepsilon_2^2}$ and $\lr{\varepsilon_2^2\delta d_{\perp}/d_{\perp}}$. The results based on event averaging via $\nqp$ are shown in the right two columns, and the results based on event averaging via $\npart$, already presented before in Figs.~\ref{fig:3},\ref{fig:7},\ref{fig:8} and \ref{fig:10}, are repeated in the left two columns.

For all observables and in almost all cases, the coefficients are quite consistent between U+U and Zr+Zr. Clear differences between event averaging based on $\npart$ and those based on $\nqp$ are also visible in the UCC region, reflecting the effects of volume fluctuations. These differences are negligible for $\lr{\varepsilon_2^2}$, but reach up to 20\% for $C_{\mathrm{d}}\{2\}$ and $\lr{\varepsilon_2^2\delta d_{\perp}/d_{\perp}}$; they are even larger for $C_{\mathrm{d}}\{3\}$, and $C_{\mathrm{d}}\{4\}$. What this means is that by selecting extremely central events, one might introduce a large bias from volume fluctuations on skewness and kurtosis. Therefore, the optimal centrality range to maximize the deformation effects, yet avoid strong volume fluctuations, should not be too narrow. A more reasonable choice would be 0\%--1\% or 0\%--5\%. In general, the magnitudes of $c'$ are much smaller than $b'$, except for skewness $C_{\mathrm{d}}\{3\}$ and $\lr{\varepsilon_2^2\delta d_{\perp}/d_{\perp}}$ in central collisions where $|c'|\gg |b'|$. The latter reinforces earlier conclusion that three-particle correlations involving $v_2$ and $[\pT]$ in heavy ion collisions are sensitive probe of the nuclear triaxiality. In some limited cases such as the $b'$ parameter for $C_{\mathrm{d}}\{3\}$ and $C_{\mathrm{d}}\{4\}$, the results are quantitatively different between the nucleon Glauber model and the quark Glauber model (compare the symbols with the lines), suggesting that the deformation contribution to high-order cumulants of $d_{\perp}$ are also sensitive to the subnucleon fluctuations.

Table~\ref{tab:3} lists the values of $a'$, $b'$ and $c'$ from Eq.~\eqref{eq:17} in the 0\%--1\% most central collisions for the four cases for calculating the observable and performing event averaging. One sees that the values of $a'$ could differ by up to a factor of 2 among the four cases. From these values, one derives the analytical function form for the $(\beta_2,\gamma)$ dependence for each observable, including various normalized cumulants discussed in pervious sections.
\begin{table}[!h]
\centering
\begin{tabular}{c|ccc|ccc|ccc|ccc}\hline
 &\multicolumn{3}{c|}{}& \multicolumn{3}{c|}{} & \multicolumn{3}{c|}{} & \multicolumn{3}{c}{}\\[-0.2ex]
variable calculation &\multicolumn{3}{c|}{nucleon}& \multicolumn{3}{c|}{quark} & \multicolumn{3}{c|}{nucleon} & \multicolumn{3}{c}{quark}\\[2ex]
event class    &\multicolumn{3}{c|}{$\npart$ }&\multicolumn{3}{c|}{$\npart$}& \multicolumn{3}{c|}{$\nqp$}& \multicolumn{3}{c}{$\nqp$}\\[2ex]\hline\hline
 &\multicolumn{3}{c|}{}& \multicolumn{3}{c|}{} & \multicolumn{3}{c|}{} & \multicolumn{3}{c}{}\\[-0.2ex]
&$a'$&$b'$&$c'$&$a'$&$b'$&$c'$&$a'$&$b'$&$c'$&$a'$&$b'$&$c'$\\[2ex]\hline
$\lr{(\frac{\delta d_{\perp}}{d_{\perp}})^2}\times10^2$&0.033&0.93&0.0039&  0.038&0.88&-0.015&  0.039&0.83&0.019&  0.04&0.85&0.023\\[2ex]
$a'+(b'+c'\cos(3\gamma))\beta_2^2$&\multicolumn{3}{c|}{}& \multicolumn{3}{c|}{} & \multicolumn{3}{c|}{} & \multicolumn{3}{c}{}\\[1ex]\hline
$\lr{(\frac{\delta d_{\perp}}{d_{\perp}})^3}\times10^4$&0.006&1.3&3.0&  0.0084&0.72&2.7&  0.012&-0.087&2.2&  0.0085&-0.43&2.4\\[2ex]
$a'+(b'+c'\cos(3\gamma))\beta_2^3$&\multicolumn{3}{c|}{}& \multicolumn{3}{c|}{} & \multicolumn{3}{c|}{} & \multicolumn{3}{c}{}\\[1ex]\hline
$(\lr{(\frac{\delta d_{\perp}}{d_{\perp}})^4}-3\lr{(\frac{\delta d_{\perp}}{d_{\perp}})^2}^2)\times10^5$&0.00033&-5.4&1.1&  0.00065&-5.0&0.88&  0.00064&-3.1&-0.1&  0.00052&-3.4&-0.35\\[2ex]
$a'+(b'+c'\cos(3\gamma))\beta_2^4$&\multicolumn{3}{c|}{}& \multicolumn{3}{c|}{} & \multicolumn{3}{c|}{} & \multicolumn{3}{c}{}\\[1ex]\hline
$\lr{\varepsilon_2^2}\times10$&0.045&2.35&0.11&  0.055&2.38&0.083&  0.047&2.32&-0.19&  0.056&2.34&-0.21\\[2ex]
$a'+(b'+c'\cos(3\gamma))\beta_2^2$&\multicolumn{3}{c|}{}& \multicolumn{3}{c|}{} & \multicolumn{3}{c|}{} & \multicolumn{3}{c}{}\\[1ex]\hline
$\lr{\varepsilon_2^2\frac{\delta d_{\perp}}{d_{\perp}}}\times10^2$&0.00051&-0.066&-1.36&  0.00070&-0.12&-1.35&  0.00097&-0.17&-1.17&  0.00084&-0.19&-1.19\\[2ex]
$a'+(b'+c'\cos(3\gamma))\beta_2^3$&\multicolumn{3}{c|}{}& \multicolumn{3}{c|}{} & \multicolumn{3}{c|}{} & \multicolumn{3}{c}{}\\[1ex]\hline
\end{tabular}
\caption{\label{tab:3} The values of the coefficients $a'$, $b'$ and $c'$ of Eq.~\eqref{eq:17} for each observable in 0--1\% U+U collisions from the Glauber model. They are listed for four cases: variables can be calculated with either nucleons or quarks and the event averaging are also based on either nucleons or quarks.} 
\end{table}

\section{Summary and a proposal}\label{sec:5}

I have shown that the two bulk quantities of the initial overlap of the heavy ion collisions, the $\varepsilon_2$ and $d_{\perp}$, which quantifies the quadrupole shape and density gradient (or the inverse size) of the overlap region, respectively, are directly related to the quadrupole deformation parameters $(\beta_2,\gamma)$ of the colliding nuclei. Aided by hydrodynamic response in the final state, these initial quantities are transformed into the experimentally measured elliptic flow $v_2$ and average transverse momentum $[\pT]$ in each event. Using an analytical argument and a Glauber model simulation, I derive analytical relations between the cumulants of $\varepsilon_2$/$d_{\perp}$ and $(\beta_2,\gamma)$. Remarkably, the variances depend mainly on $\beta_2$ (i.e. $\lr{\varepsilon_2^2}, \lr{(\delta d_{\perp}/d_{\perp})^2}\sim a'+b'\beta_2^2$), while the skewness are sensitive to both parameters in a simple factorizable form (i.e. $\lr{\varepsilon_2^2 \delta d_{\perp}/d_{\perp}}, \lr{(\delta d_{\perp}/d_{\perp})^3}\sim  a'+(b'+c'\cos(3\gamma))\beta_2^3$). Similar analytical relations are naturally expected for final-state observables involving $v_2$ and $[\pT]$. These robust relations provide an efficient way,  via a dedicated system scan, to constrain simultaneously the $\beta_2$ and $\gamma$ of the atomic nuclei. 
\begin{figure}[h!]
\begin{center}
\includegraphics[width=0.8\linewidth]{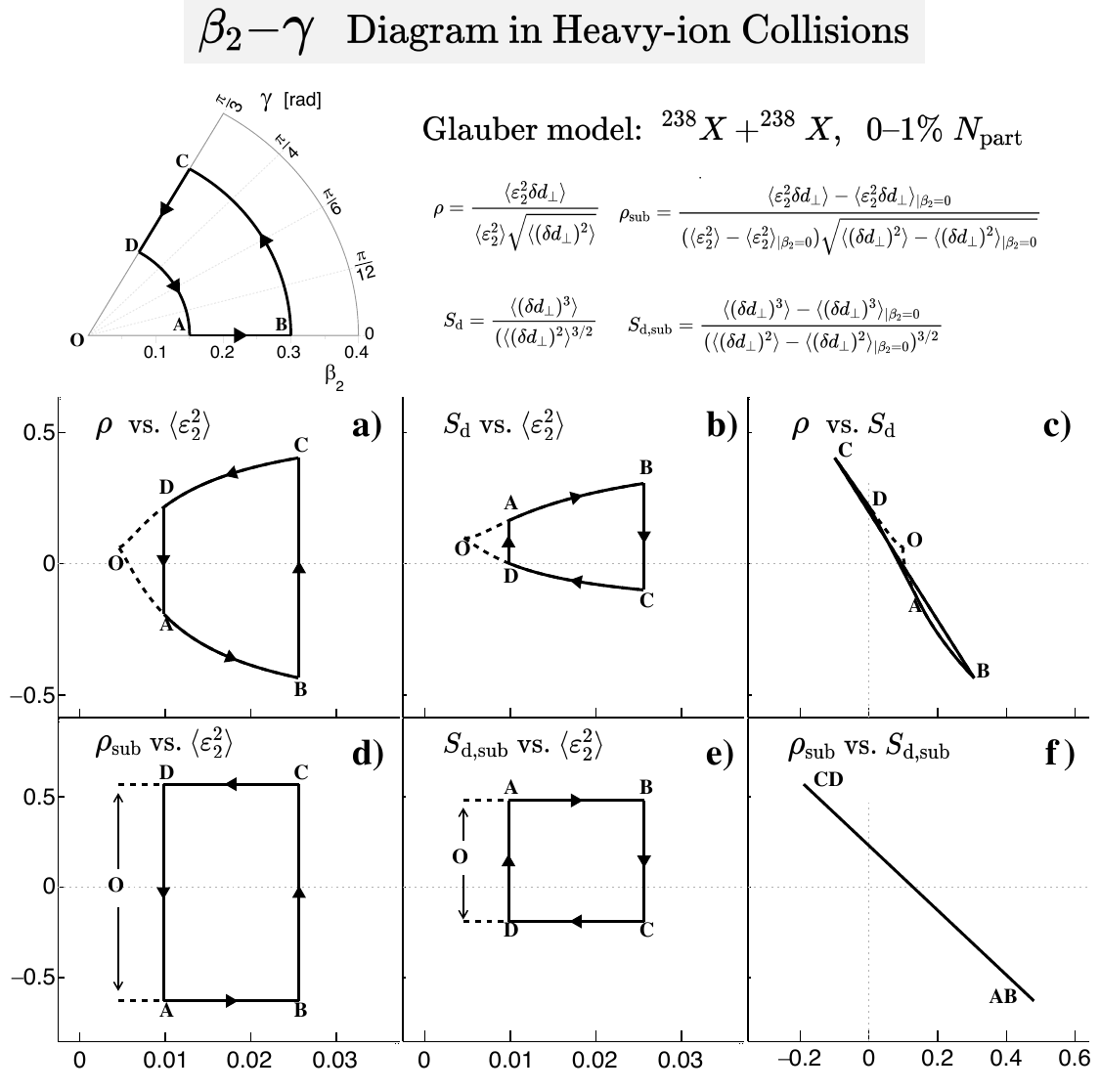}
\end{center}\vspace*{-0.5cm}
\caption{\label{fig:17} Glauber model prediction of the mapping of a closed trajectory on the $(\beta_2,\gamma)$ plane from nuclear structure side (top-left) onto a trajectory on the ($\rho$, $\langle\varepsilon_2^2\rangle$) plane (panel-a), the ($S_{\mathrm{d}}$,$\langle\varepsilon_2^2\rangle$) plane (panel-b), as well as that for the baseline-subtracted quantity ($\rho_{\mathrm{sub}}$, $\langle\varepsilon_2^2\rangle$) (panel-d) and ($S_{\mathrm{d,sub}}$, $\langle\varepsilon_2^2\rangle$) (panel-e). The definition of these quantities are given in the top-right corner. The trajectories are also shown on the ($\rho$, $S_{\mathrm{d}}$) plane (panel-c) and on the ($\rho_{\mathrm{sub}}$,$S_{\mathrm{d,sub}}$) plane (panel-f). The results are shown for collision of nucleus with 238 nucleons and for the 0\%--1\% most central events selected based on $\npart$. Note that the correlation with variance $\langle(\delta d_{\perp}/d_{\perp})^2\rangle$ as the $x$-axis are similar, i.e. only require a shift and rescaling (see text).}
\end{figure}

To illustrate how this can be done, one refers to the results obtained from Glauber model for 0\%--1\% most central U+U collisions from the first column of Tab.~\ref{tab:3},
\begin{align}\nonumber
\langle\varepsilon_2^2\rangle&\approx[0.02+\beta_2^2]\times0.235\\\nonumber
\langle(\delta d_{\perp}/d_{\perp})^2\rangle&\approx[0.035+\beta_2^2]\times 0.0093\\\nonumber
\langle(\delta d_{\perp}/d_{\perp})^3\rangle&\approx[0.006+(1.3+3.0\cos(3\gamma))\beta_2^3]\times10^{-4}\\\label{eq:26}
\langle\varepsilon_2^2\delta d_{\perp}/d_{\perp}\rangle&\approx[0.0005-(0.07+1.36\cos(3\gamma))\beta_2^3]\times10^{-2}
\end{align}
From these I construct ratios $\rho(\varepsilon_2^2,\delta d_{\perp}/d_{\perp})$ and $S_{\mathrm{d}}$, as well as baseline subtracted ratios $\rho_{\mathrm{sub}}$ and $S_{\mathrm{d,sub}}$ (their definitions are repeated in Fig.~\ref{fig:17}). Eq.~\eqref{eq:26} can map any trajectory in the $(\beta_2,\gamma)$ diagram from low-energy nuclear structure side (so-called ``Hill-Wheeler'' coordinate) onto new trajectories in various correlation plots from high-energy side as shown in the bottom panels (a)--(f).  I note that the direction of the trajectory in the $(\rho,\langle\varepsilon_2^2\rangle)$ plane is opposite to that in the $(S_{\mathrm{d}},\langle\varepsilon_2^2\rangle)$ plane, and the trajectory in the $(\rho,S_{\mathrm{d}})$ plane almost collapses into a straight line. The $\gamma$ dependences in these plots follow a simple linear function of $\cos(3\gamma)$, while the $\beta_2$ dependence is more complex due to the offsets in Eq.~\eqref{eq:26}. The correlations are much well behaved for $\rho_{\mathrm{sub}}$ and $S_{\mathrm{d,sub}}$ as shown in the bottom row of Fig.~\ref{fig:17}. In particular, the differences between prolate and oblate deformation for these quantities are independent of $\beta_2$, and they are also expected to be nearly independent of centrality as suggested by Figs.~\ref{fig:9} and \ref{fig:11}. Therefore, one could determine the $\gamma$ angle of any nucleus with similar mass number, once the values of $\rho_{\mathrm{sub}}$ and $S_{\mathrm{sub}}$ are calibrated from collisions of prolate and oblate nuclei with known $\beta_2$.

A few additional summarizing points can be made about these flow diagrams. 1) One can replace the $x$-axis with $\langle(\delta d_{\perp}/d_{\perp})^2\rangle$, the trajectories would be shifted and rescaled but their shapes remain the same.  2) Since the coefficients $b'$ and $c'$ are relatively insensitive to the size of the collision systems, the correlations in the bottom row of Fig.~\ref{fig:17} are expected to be valid for all medium and large nuclei. By the way, the change of $\rho_{\mathrm{sub}}$ and $S_{\mathrm{d,sub}}$ when nuclear shape is varied from prolate to oblate, unlike $\rho$ and $S_{\mathrm{d}}$, are also relatively independent of centrality. This implies that the curves in the bottom panels only shift vertically and narrow horizontally for events in mid-central collisions, but the height remains roughly the same. 3) One should be able to construct similar flow diagrams for cumulants of $v_2$ and $[\pT]$ in the final state. This can be estimated from the well-known linear relation $v_2\propto\varepsilon_2$ and $\delta [\pT]/[\pT]\propto \delta d_{\perp}/d_{\perp}$, or more precisely evaluated from the full hydrodynamic model simulations. 4) The generalization of this idea to kurtosis and higher-order cumulants may not work well due to strong  nonlinear mode mixing from lower-order cumulants. 

Study of the nuclear deformation, in particular shape evolution in the $(\beta_2,\gamma)$ diagram along the isobaric chain by adding neutron and protons, is one of the most important areas of research in nuclear structure community~\cite{Heyde2011}. High-energy heavy-ion collisions offer a new tool to image the shape of atomic nuclei by smashing them together and measure the collective flow response in the final state. The skewness $\langle(\delta d_{\perp}/d_{\perp})^3\rangle$ and $\langle\varepsilon_2^2\delta d_{\perp}/d_{\perp}\rangle$, experimentally accessible via three-particle correlations $\langle(\delta [p_{\mathrm{T}}]/[p_{\mathrm{T}}])^3\rangle$ and $\langle v_2^2\delta [p_{\mathrm{T}}]/[p_{\mathrm{T}}]\rangle$, show remarkably strong sensitivity to triaxiality over a broad range of centrality, as well as nearly system-size independent signal strength. The existing data from various species, in particular the recent isobar $^{96}$Zr+$^{96}$Zr and $^{96}$Ru+$^{96}$Ru collision data~\cite{STAR:2021mii} at high energy, provide a unique opportunity to test the methodology proposed in this paper~\cite{Jia:2021oyt,Zhang:2021kxj}. However, most valuable information will ultimately arise from a collision scan of systems for which one already have precision knowledge from the nuclear structure community to calibration the hydrodynamic response, followed by application to systems for which one not have sufficient understanding. 

{\bf Acknowledgements:} I am grateful for the AMPT simulation data provided by Chunjian Zhang. I thank Giuliano Giacalone, Chunjian Zhang and Somadutta Bhatta for valuable discussions. This work is supported by DOE DEFG0287ER40331.

\appendix

\section{AMPT model}\label{sec:app0}
I have shown that the initial state of the heavy ion collisions are very sensitive to quadrupole deformation and triaxiality of the colliding nuclei, and I have constructed multiple observables to constrain $\beta_2$ and $\gamma$ independently.  The next crucial question, however, is how much of these sensitivities in the initial state survive to the particle correlations in the final state. Previous hydrodynamic model studies and data comparisons have firmly established the proportionality between $\varepsilon_2$ and $v_2$, and to lesser extent also the positive correlation between $d_{\perp}$ and $[\pT]$~\cite{Bozek:2012fw,Bozek:2017jog} and between $\lr{\varepsilon_2^2,\delta d_{\perp}}$ and $\lr{v_2^2,\delta [\pT]}$~\cite{Schenke:2020uqq,Giacalone:2020dln}. 

To understand the conversion from $\varepsilon_2$ and $d_{\perp}$ in the initial overlap to $v_2$ and $[\pT]$ in the final state, the popular event generator ``a multi-phase transport model'' (AMPT)~\cite{Lin:2004en} is used, which is a realistic yet computationally efficient way to implement hydrodynamic response.  The AMPT model has been demonstrated to qualitatively describe the harmonic flow $v_n$ in $p$+A and A+A collisions~\cite{Xu:2011jm,Xu:2011fe}, so it can be use to predict the $\beta_2$ dependence of $v_n$. A previous study has demonstrated a robust simple quadratic dependence $\lr{v_2^2}=a+b\beta_2^2$ in the final state as a result of a linear response to a similar dependence in the initial state $\lr{\varepsilon_2^2}=a'+b'\beta_2^2$~\cite{Giacalone:2021udy,Jia:2021wbq}. However this model is known to have the wrong hydrodynamic response for the radial flow, i.e. the centrality dependence of average transverse momentum $\llrr{\pT}\equiv\lr{[\pT]}$ and the variance $\lr{(\delta [\pT])^2}$ do not describe the experimental data~\cite{Ma:2016fve,Jia:2021wbq}. A recent modification of the model~\cite{Zhang:2021vvp} fixed the problem with the $\llrr{\pT}$, but the value of $\lr{(\delta [\pT])^2}$ is still more than a factor of 3 lower than the STAR data~\cite{Adam:2019rsf,jjia}~\footnote{Hydrodynamic model simulation based on Trento initial condition~\cite{Giacalone:2020lbm} predicts a much larger $[\pT]$ fluctuation, but with very little sensitivity on $\beta_2$.}. This implies that the response of $[\pT]$ to $d_{\perp}$ in AMPT is a lot weaker than the experimental finding, and explains why the model fail to describe quantitatively the behavior of $\lr{v_2^2\delta[\pT]}$ in U+U collisions observed in the STAR data~\cite{jjia}. Nevertheless, since the response of $v_2$ is correct, one can still study the parametric $(\beta_2,\gamma)$ dependence of $\lr{v_2^2\delta[\pT]}$ and compare with the trend of $\lr{\varepsilon_2^2\delta d_{\perp}}$. However, this unfortunately can not be said about cumulants of $[\pT]$ fluctuations.

Following Refs~\cite{Ma:2014pva,Bzdak:2014dia,Nie:2018xog}, I use the AMPT model v2.26t5 with string-melting mode and partonic cross section of 3.0~mb, which I check reasonably reproduce Au+Au $v_2$ data at RHIC. The Woods-Saxon parameters in the AMPT are chosen to be $R_0=6.81$fm and $a=0.54$ similar to ~\cite{Heinz:2004ir} but with different fixed values of $(\beta_2, \gamma)$. The $v_2$ and $[\pT]$ are calculated with all hadrons with $0.2<\pT<2$ GeV and $|\eta|<2$, and the event centrality is defined using either $\npart$ or inclusive hadron multiplicity in $|\eta|<2$, $\nall$. The value of $\nall$, which include both charged and neutral particles, is about six times of the charged hadron multiplicity density, i.e. $\nall\approx 6 dN_{\mathrm{ch}}/d\eta$.

\begin{figure}[h!]
\begin{center}
\includegraphics[width=0.4\linewidth]{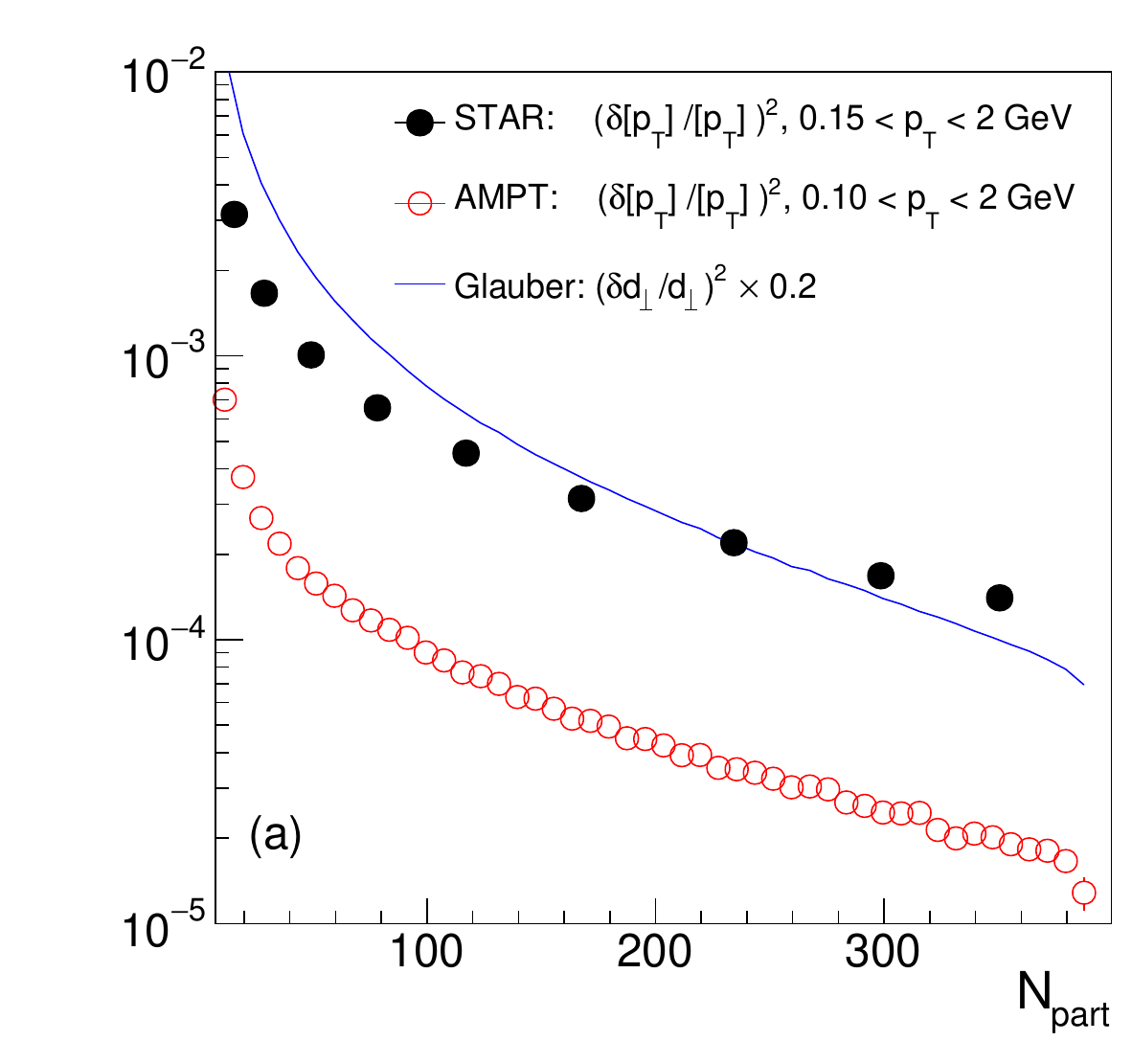}\includegraphics[width=0.4\linewidth]{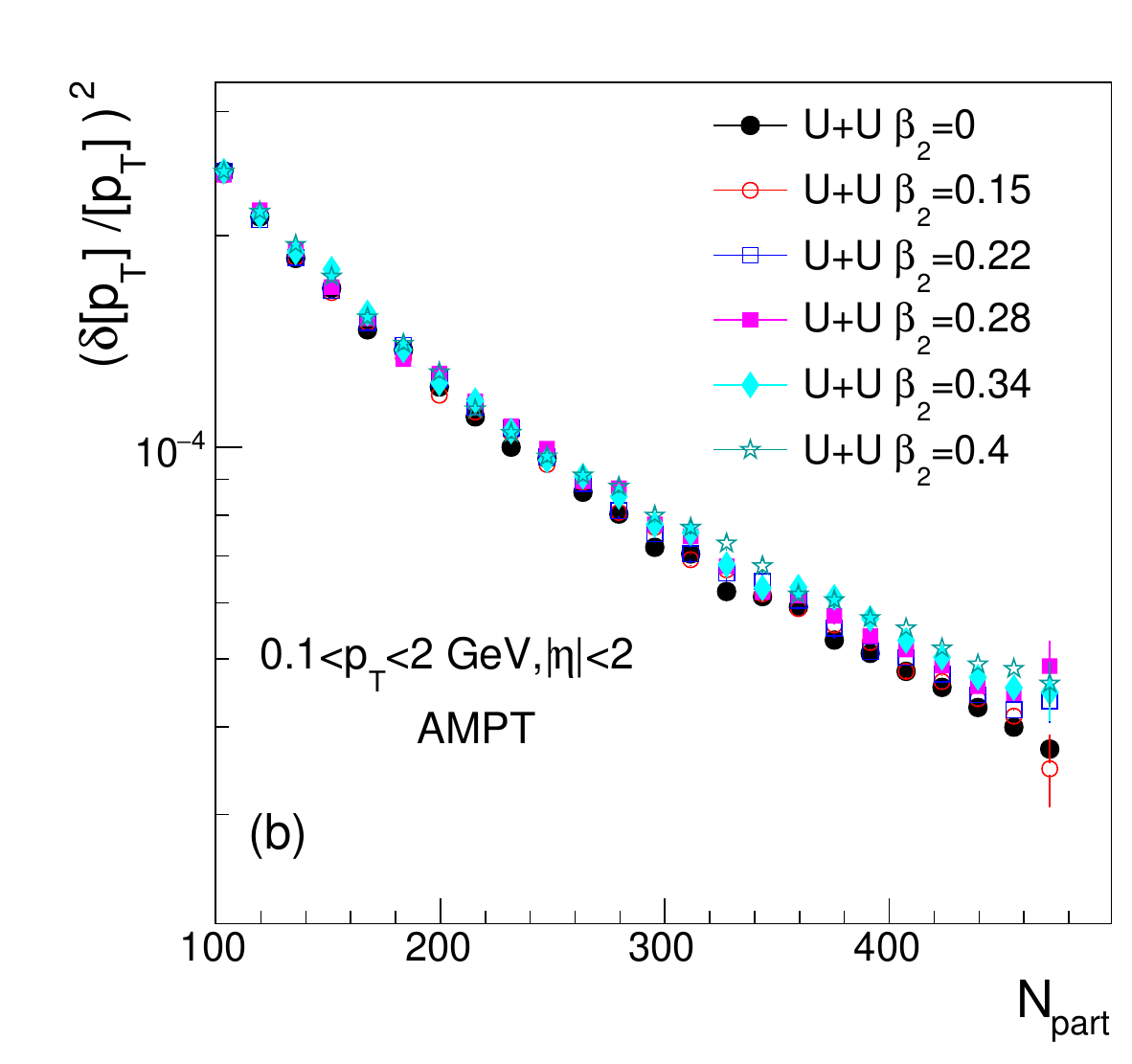}
\end{center}\vspace*{-0.3cm}
\caption{\label{fig:14} Left: variance of $[\pT]$ fluctuation from AMPT model (open symbol) and experimental data Ref.~\cite{Adam:2019rsf} (solid symbol), as well as the variance of $d_{\perp}$ (solid line) in Au+Au collisions at $\sqrtsnn=200$ GeV. Right: variance of $[\pT]$ from AMPT model in U+U collisions for different values of $\beta_2$. }
\end{figure}

One main drawback of the AMPT model is that it underestimates the hydrodynamic response of radial flow. For one thing, it undershoots the variance of the $\pT$ fluctuations from data, see the left panel of Fig.~\ref{fig:14}. The right panel show that the AMPT model predicts a very weak dependence of $\lr{(\delta [\pT]/[\pT])^2}$ on $\beta_2$. Even for a value of $\beta_2=0.28$, the increase of $[\pT]$ variance is only 30\%. Similar observation is also made for  $\lr{(\delta [\pT]/[\pT])^3}$ (not shown). This is in clear contradiction to the much larger influence from deformation observed in the recent experimental results of variance and skewness of $[\pT]$ in U+U and Au+Au collisions~\cite{jjia}. Hence, AMPT model can not be used to study reliably the deformation effects on the $[\pT]$ fluctuations. Instead, I shall focus on $\lr{v_2^2\delta [\pT]}$, the rationale being that even though the radial flow response is underestimated, the elliptic flow response is still correctly modeled. I hope to at least explore the qualitative features of $\lr{v_2^2\delta [\pT]}$ and compare to $\lr{\varepsilon_2^2\delta d_{\perp}}$. 

\begin{figure}[h!]
\begin{center}
\includegraphics[width=0.9\linewidth]{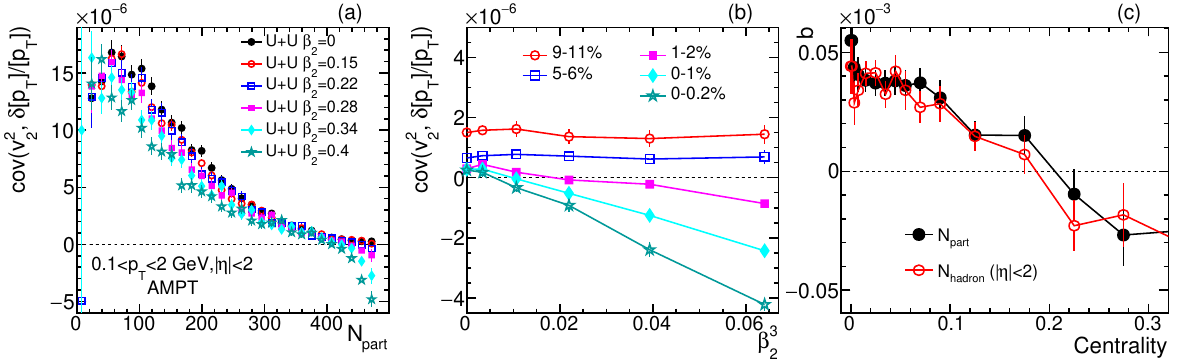}\\
\includegraphics[width=0.9\linewidth]{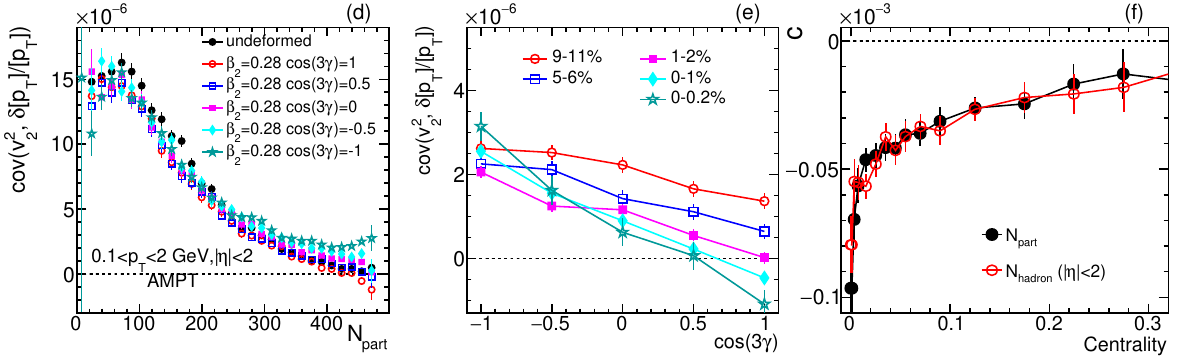}
\end{center}\vspace*{-0.5cm}
\caption{\label{fig:15}  The $\lr{v_2^2\delta [\pT]/[\pT]}$ for several $\beta_2$ values of prolate shape $\gamma=0$ (top row) and several $\gamma$ values with $\beta_2=0.28$ (bottom row) in U+U collisions from the AMPT model. The left column show the $\npart$ dependence. The middle column shows the results as a function of $\beta_2^3$ (top panel) or $\cos(3\gamma)$ (bottom panel) in several centrality ranges based on $\npart$. The right column summarizes the coefficients $b$ (top) and $c$ (bottom) from Eq.~\eqref{eq:24} as a function of centrality based on $\npart$ (filled symbols) or $\nall$ (open symbols).}.
\end{figure}

The left column of Fig.~\ref{fig:15} shows the $\npart$ dependence of $\lr{v_2^2\delta [\pT]/[\pT]}$ for several values of $\beta_2$ and $\gamma$, calculated using the multi-particle correlation framework of Ref.~\cite{Zhang:2021phk}. There are clear sensitivity on both parameters, especially in the UCC region. The values are integrated over several centrality ranges and plotted as a function of $\beta_2^3$ and $\cos(3\gamma)$ in the middle column, calculated from the corresponding data in the left column. Despite the large statistical uncertainties, linear dependences are observed, confirming the trends seen in the Glauber model:
\begin{align}\label{eq:24}
\lr{v_2^2(\delta [\pT]/[\pT])} = a+(b+c\cos(3\gamma)) \beta_2^3\;. 
\end{align}
The values of $b$ and $c$ are shown in the right column as a function of centrality; the centrality-dependent trends are similar to those obtained from Glauber model (compare to Fig.~\ref{fig:10}). However, the values of $b$ and $c$ are about a factor of 100 smaller than $b'$ and $c'$, also $b$ is larger than 0 in central collisions, while $b'$ is less than 0 over the full centrality range. In hydrodynamic model with linear response assumption of Eq.~\eqref{eq:3}, one has approximately,
\begin{align}\label{eq:25}
\lr{v_2^2\frac{\delta [\pT]}{[\pT]}} \approx k_2^2 k_0\lr{\varepsilon_2^2 \frac{\delta d_{\perp}}{d_{\perp}}}
\end{align}
Using the value of $k_2\approx0.2$ from a hydrodynamic model~\cite{Song:2010mg} and $k_0\approx0.4$ from left panel of Fig.~\ref{fig:11} in central collisions, one expects a factor of 60. I also repeat the same analysis using $\nall$ to classify events. They give very similar values of $b$ and $c$ as shown in the right column of Fig.~\ref{fig:15}, implying the results are robust against the volume fluctuations. 

\begin{figure}[h!]
\begin{center}
\includegraphics[width=0.5\linewidth]{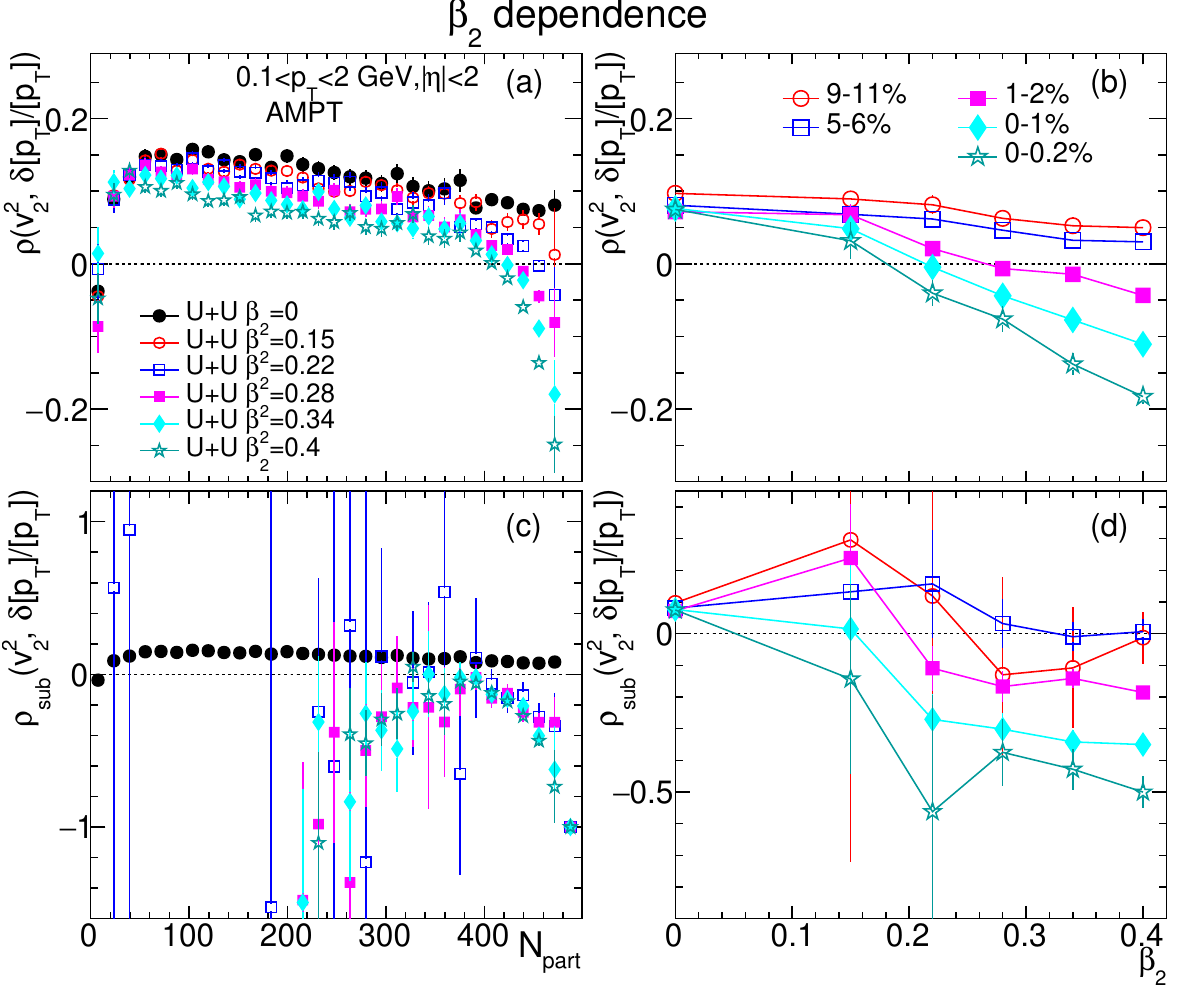}\includegraphics[width=0.5\linewidth]{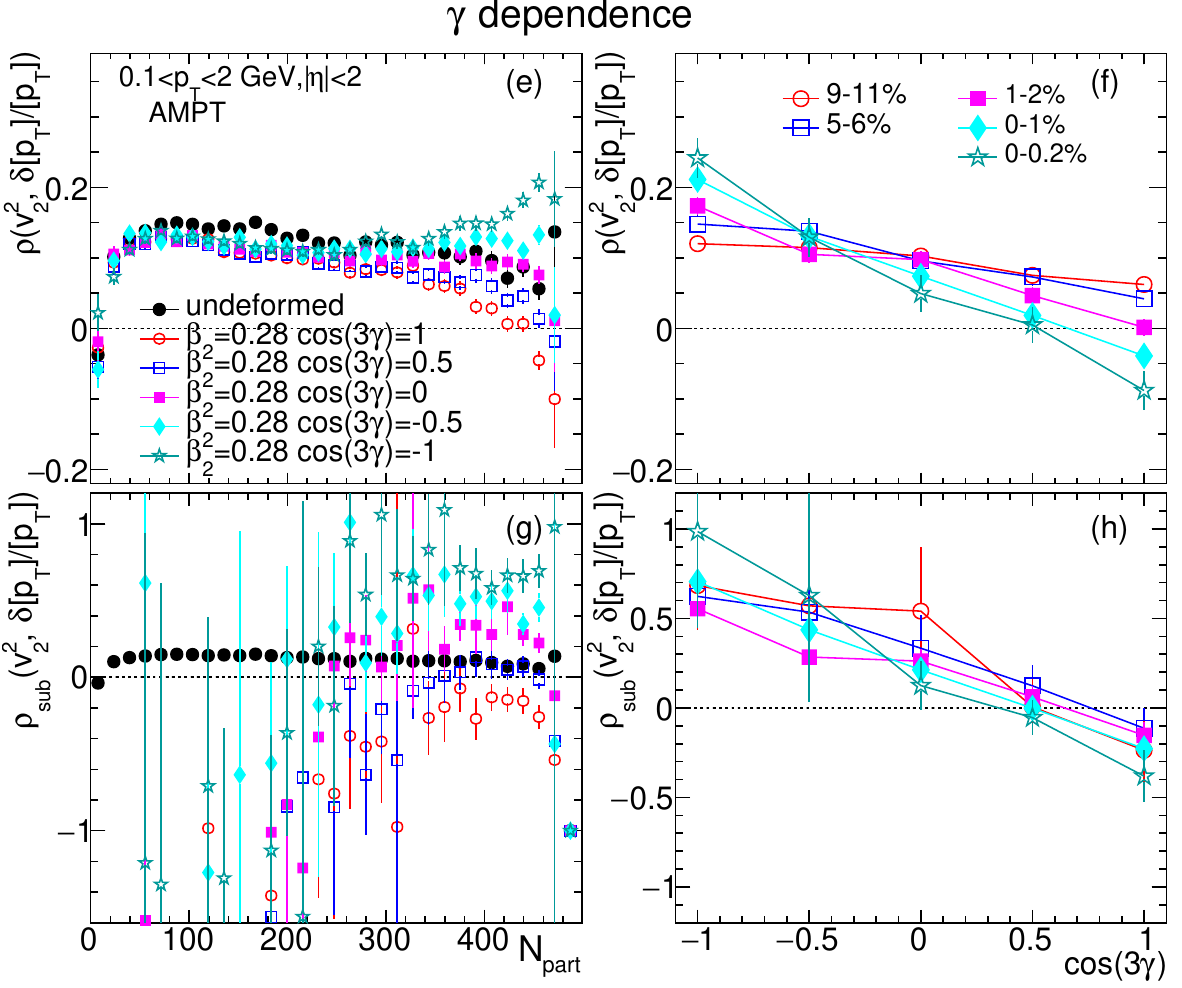}
\end{center}\vspace*{-0.5cm}
\caption{\label{fig:16}  Left Part: The $\rho(v_2^2,\delta [\pT]/[\pT])$ (top row) and $\rho_{\rm{sub}}(v_2^2,\delta [\pT]/[\pT])$ (bottom row) as a function of $\npart$ for several $\beta_2$ values of prolate shape $\gamma=0$ (left column) and as a function of $\beta_2$ in several centrality ranges based on $\npart$ (right column). Right part: The $\rho(v_2^2,\delta [\pT]/[\pT])$ (top row) and $\rho_{\rm{sub}}(v_2^2,\delta [\pT]/[\pT])$ (bottom row) as a function of $\npart$ for several $\gamma$ values with $\beta_2=0.28$ (left column) and as a function of $\cos(3\gamma)$ in several centrality ranges based on $\npart$ (right column).}
\end{figure}

From these results, I calculate the normalized quantities, $\rho(v_2^2,\frac{\delta [\pT]}{[\pT]})$ and $\rho_{\mathrm{sub}}(v_2^2,\frac{\delta [\pT]}{[\pT]})$, defined similar to those in Eqs.~\eqref{eq:14a} and \eqref{eq:21}. The results are shown in Fig.~\ref{fig:16} for $\beta_2$ dependence on the left part and $\gamma$ dependence on the right part. The $\rho$ follows approximately a linear dependence of $\beta_2$, similar to Glauber model results (top panel in the second column of Fig.~\ref{fig:11}). The $\rho_{\mathrm{sub}}$ in the bottom panels are nearly independent of $\beta_2$ as expected. For the $\cos(3\gamma)$ dependence, $\rho$ data exhibit different slopes for different centralities ranges, but $\rho_{\mathrm{sub}}$ data follow a common slope in all centrality ranges. What this means is that the difference of $\rho_{\mathrm{sub}}$ between prolate and oblate is approximately independent of centrality, similar to the results from Glauber model shown in the bottom right panel of Fig.~\ref{fig:11}.

\section{Transverse size fluctuations in head-on collisions}\label{sec:app1}

Although the shape and size of atomic nuclei with static deformation is fixed in the intrinsic frame, the shape and size for the distribution projected to the transverse plane $(x,y)$ in the laboratory frame depend on the Euler angle $\Omega$, and therefore fluctuates event to event. The expression for ${\bm \epsilon}_{2}$ has been derived in the Appendix A of Ref.~\cite{Jia:2021tzt}, I shall focus on $\delta d_{\perp}/d_{\perp}$ in Eq.~\eqref{eq:11}.

First, I express the angular weights of variance and covariance of the coordinates, $\lr{x^2}=\lr{r^2\sin^2\theta\sin^2\phi}$, $\lr{y^2}=\lr{r^2\sin^2\theta\cos^2\phi}$ and $\lr{xy}=\lr{r^2\sin^2\theta\sin\phi\cos\phi}$ in terms of spherical harmonics,
\begin{align}\nonumber
&\sin^2\theta\cos^2\phi = \frac{1}{3}+ \sqrt{\frac{2\pi}{15}}(Y_2^2+Y_2^{-2})-\frac{2}{3}\sqrt{\frac{\pi}{5}}Y_2^0\;,\;\sin^2\theta\sin^2\phi = \frac{1}{3}- \sqrt{\frac{2\pi}{15}}(Y_2^2+Y_2^{-2})-\frac{2}{3}\sqrt{\frac{\pi}{5}}Y_2^0\;,\\\label{eq:app1}
&\sin^2\theta\sin\phi\cos\phi=-i\sqrt{\frac{2\pi}{15}}(Y_2^2-Y_2^{-2}).
\end{align}
In the rotated frame, one needs to apply the substitution $Y_2^m\rightarrow \sum_{m'} D_{m,m'}^2(\Omega) Y_2^{m'}$. Keeping the leading order term in $\beta_2$, using the notation $\alpha_0=\cos\gamma$, $\alpha_2=\alpha_{-2}=\sin\gamma/\sqrt{2}$ for quadrupole deformation, the variances and covariance become
\begin{align}\nonumber
\lr{x^2,y^2} &= \frac{\int \rho(r)r^4dr \int (1+\beta_2 \sum_m \alpha_{m} Y_2^m)^5 [\frac{1}{3}\pm\sqrt{\frac{2\pi}{15}} \sum_{m'}(D_{2,m'}^2+D_{-2,m'}^2)Y_2^{m'}-\frac{2}{3}\sqrt{\frac{\pi}{5}}\sum_{m'}D_{0,m'}^2 Y_2^{m'}]\sin\theta d\theta d\phi}{\int \rho(r)r^2dr\int (1+\beta_2 \sum_m \alpha_{m} Y_2^m)^3 \sin\theta d\theta d\phi}\\\label{eq:app2}
&\approx \frac{R_0^2}{5}\left[1+\sqrt{\frac{5}{4\pi}}\beta_2\sum_{m'}\alpha_{m'}\left(-\sum_{m'}D_{0,m'}^2\pm \sqrt{\frac{3}{2}}(D_{2,m'}^2+D_{-2,m'}^2)\right)\right]\\\label{eq:app3}
\lr{xy} &\approx -i\frac{R_0^2}{5}\frac{15\beta_2}{4\pi}\sum_{m'}\alpha_{m'}(D_{2,m'}^2-D_{-2,m'}^2)\;.
\end{align}
 The transverse area $S_{\perp}$ in the projected plane has the following expression
\begin{align}\label{eq:app4}
\frac{S_{\perp}^2}{\pi^2} = \lr{x^2}\lr{y^2}-\lr{xy}^2 = \frac{R_0^4}{25}\left[1-\sqrt{\frac{5}{\pi}}\beta_2\sum_{m}\alpha_{m}D_{0,m}^2+\frac{5}{4\pi}\beta_2^2\sum_{m,m'}\alpha_{m}\alpha_{m'}(D_{0,m}^2D_{0,m'}^2-6D_{2,m}^2D_{-2,m'}^2)\right] 
\end{align}
Keeping the leading term $\beta_2$, the fluctuation relative to the averaging over the $\Omega$ is
\begin{align}\label{eq:app5}
\frac{\delta d_{\perp}}{d_{\perp}} = -\frac{1}{4}\frac{\delta S_{\perp}^2}{S_{\perp}^2} = \sqrt{\frac{5}{16\pi}}\beta_2\sum_{m}\alpha_{m}D_{0,m}^2=\sqrt{\frac{5}{16 \pi}} \beta_{2}\left(\cos \gamma D_{0,0}^{2}+\frac{\sin \gamma}{\sqrt{2}}\left[D_{0,2}^{2}+D_{0,-2}^{2}\right]\right)\;,
\end{align}
where I have used the relation $d_{\perp}=\sqrt{\npart/S_{\perp}}$ and assumed $\npart$ is a constant in head-on collisions. 

Two comments are in order. First, the transverse area can also be defined as $S_{\perp}=\pi\left(\lr{x^2}+\lr{y^2}\right)$.  This definition gives exactly the same expression for $\delta d_{\perp}/d_{\perp}$ in the leading order of $\beta_2$. Second, in general the next-leading order contribution to $d_{\perp}$ contains terms that scale like $\beta_2^2 (\sum_m\alpha_{m}D_{0,m}^2)^2$ or $\beta_2^2 (\sum_m\alpha_{m}D_{2,m}^2)(\sum_m\alpha_{m}D_{2,m}^2)^*$. In the calculation of variances, they will appear as
\begin{align}\nonumber
\lr{(\delta d_{\perp}/d_{\perp})^2} &= \frac{5}{16\pi}\lr{\left(\beta_2\Sigma + c_1\beta_2^2 \Sigma^2+c_2\beta_2^2 \Pi^2+\mathcal{O}(\beta_2^3)\right)^2}= \frac{5}{16\pi}\beta_2^2\left[\lr{\Sigma^2}+2c_1\beta_2\lr{\Sigma^3}+2c_2\beta_2\lr{\Sigma\Pi^2}+\mathcal{O}(\beta_2^2)\right]\\\label{eq:app6}
&=\frac{1}{16\pi}\beta_2^2\left(1+\frac{4}{7}(c_1-c_2)\beta_2\cos(3\gamma)+\mathcal{O}(\beta_2^2)\right)
\end{align}
where I denote $\Sigma\equiv\sum_m\alpha_{m}D_{0,m}^2$ and $\Pi^2\equiv (\sum_m\alpha_{m}D_{2,m}^2)(\sum_m\alpha_{m}D_{2,m}^2)^*$, and the values of $c_1$ and $c_2$ depend on the definition of $d_{\perp}$. For the case in Eq.~\eqref{eq:app4}, one can show $c_2=\frac{2}{3} c_1=\frac{3}{2}\sqrt{\frac{5}{16\pi}}=0.473$. The two higher-order terms in this expansion have the same form as those in Eq.~\eqref{eq:12}, and their contributions are proportional to $\cos(3\gamma)$. They are responsible for the clear residual dependence on the triaxiality of $\lr{(\delta d_{\perp}/d_{\perp})^2}$ in Fig.~\ref{fig:5} and $\lr{\varepsilon_2^2}$ in Ref.~\cite{Jia:2021tzt}.  That is why the prolate deformation with $\beta_2=0.28$ in the left panels of Fig.~\ref{fig:5} has a smaller $\lr{(\delta d_{\perp}/d_{\perp})^2}$ value by about $8/7(c_1-c_2)\beta_2 =7\%$ in central collisions. 

Following Eq.~\eqref{eq:app6}, one can also estimate the higher-order correction to the skewness and kurtosis
\begin{align}\nonumber
\lr{(\delta d_{\perp}/d_{\perp})^3} &= (\frac{5}{16\pi})^{3/2}\lr{\left(\beta_2\Sigma + c_1\beta_2^2 \Sigma^2+c_2\beta_2^2 \Pi^2\right)^3}= (\frac{5}{16\pi})^{3/2}\beta_2^3\left[\lr{\Sigma^3}+3c_1\beta_2\lr{\Sigma^4}+3c_2\beta_2\lr{\Sigma^2\Pi^2}\right]\\\label{eq:app7}
&=\frac{\sqrt{5}}{224 \pi^{3/2}}\beta_2^3\left(\cos(3\gamma)+\frac{9c_1+3c_2}{2}\beta_2\right)
\end{align}
\begin{align}\nonumber
\lr{(\delta d_{\perp}/d_{\perp})^4}-3\lr{(\delta d_{\perp}/d_{\perp})^2}^2 &=\frac{5}{16\pi}\beta_2^4\left[\lr{\Sigma^4}-3\lr{\Sigma^2}^2+4\beta_2\left(c_1\lr{\Sigma^5}+c_2\lr{\Sigma^3\Pi^2}-3c_1\lr{\Sigma^2}\lr{\Sigma^3}-3c_2\lr{\Sigma^2}\lr{\Sigma\Pi^2}\right)\right]\\\label{eq:app8}
&=\frac{3}{896 \pi^{2}}\beta_2^4\left(-1+\frac{4}{33}(17c_1+23c_2)\beta_2\cos(3\gamma)\right)
\end{align}
 For skewness, the higher-order term leads to a positive shift for $S_{\mathrm{d}}$. For $\beta_2=0.28$, it is $\Delta S_{\mathrm{d}}/|S_{\mathrm{d}}|=\frac{9c_1+3c_2}{2} \beta_2 = 2.2$, i.e. the amount of shift is comparable to the variation from prolate and oblate deformation. In reality, one observe the shift is about 1/3 of the predicted size (see bottom-left panel of Fig.~\ref{fig:7}).  For kurtosis, the contribution is about $\Delta K_{\mathrm{d}}/|K_{\mathrm{d}}|=\frac{4}{33}(17c_1+33c_2)\beta_2 \cos3\gamma \approx 0.79 \cos3\gamma$. Assuming $K_{\mathrm{d}}=-3/7$ from Tab.~\ref{tab:2}, then $\Delta K_{\mathrm{d}} = 0.33 \cos3\gamma$, which is about a factor of 3 of what is observed in the Glauber model (top-right panel of Fig.~\ref{fig:9}).

\section{Additional plots}\label{sec:app2}

\begin{figure}[h!]
\begin{center}
\includegraphics[width=0.9\linewidth]{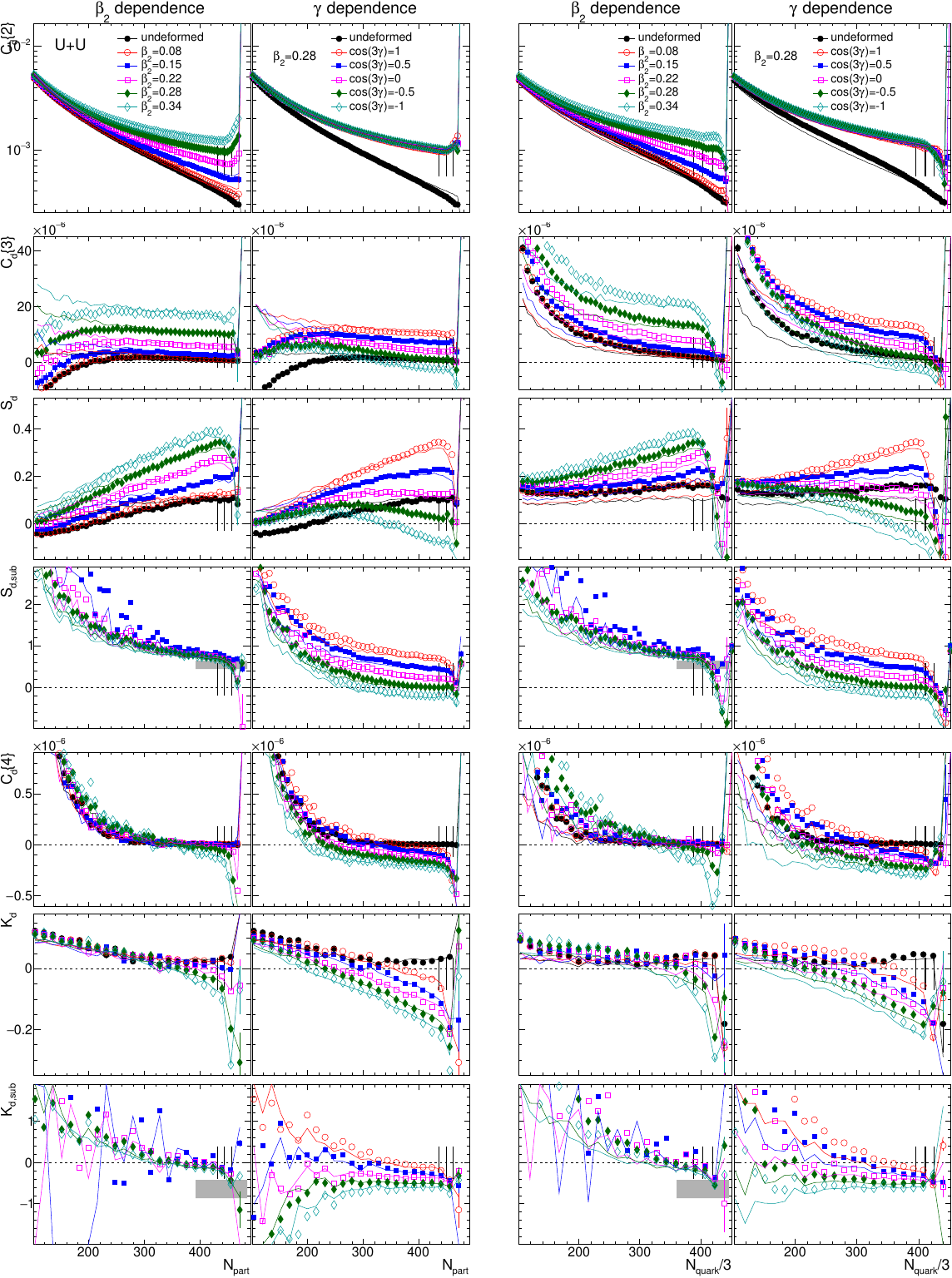}
\end{center}
\caption{\label{fig:app1} $\npart$ (left two columns) and $\nqp$ (right two columns) dependences of various $d_{\perp}$ observables as indicated by the $y$-axis title in the left side, in U+U collisions compared between different $\beta_2$ (1st and 3rd columns) and different $\gamma$ with $\beta_2=0.28$ (2nd and 4th columns), calculated with nucleons (symbols) and quarks (lines). The vertical lines in each panel correspond to locations for 0.2\%, 1\% and 2\% centralities.}
\end{figure}

\begin{figure}[h!]
\begin{center}
\includegraphics[width=0.9\linewidth]{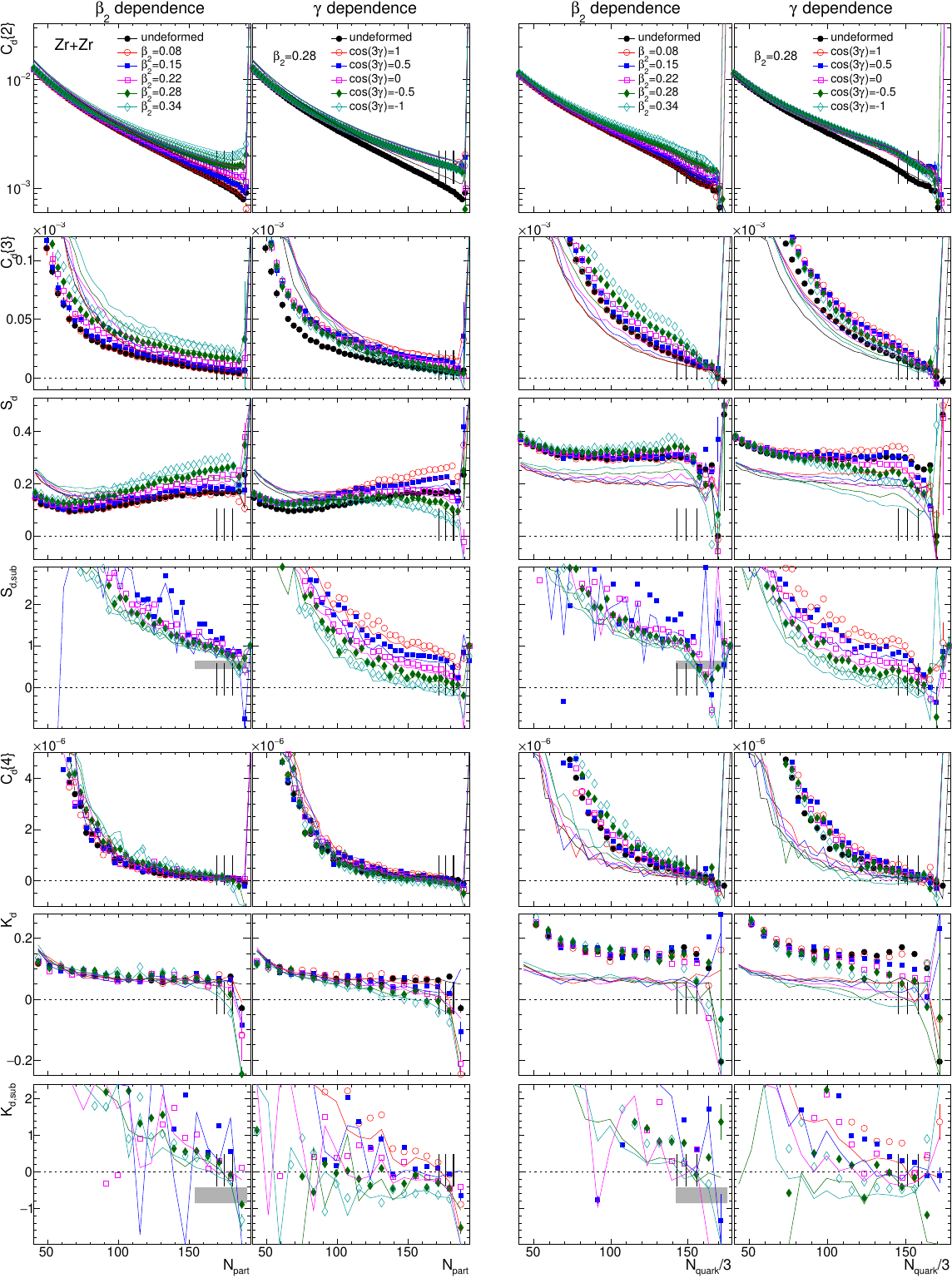}
\end{center}
\caption{\label{fig:app2} Same as Fig.~\ref{fig:app1} but for Zr+Zr collisions. }
\end{figure}

\begin{figure}[h!]
\begin{center}
\includegraphics[width=0.99\linewidth]{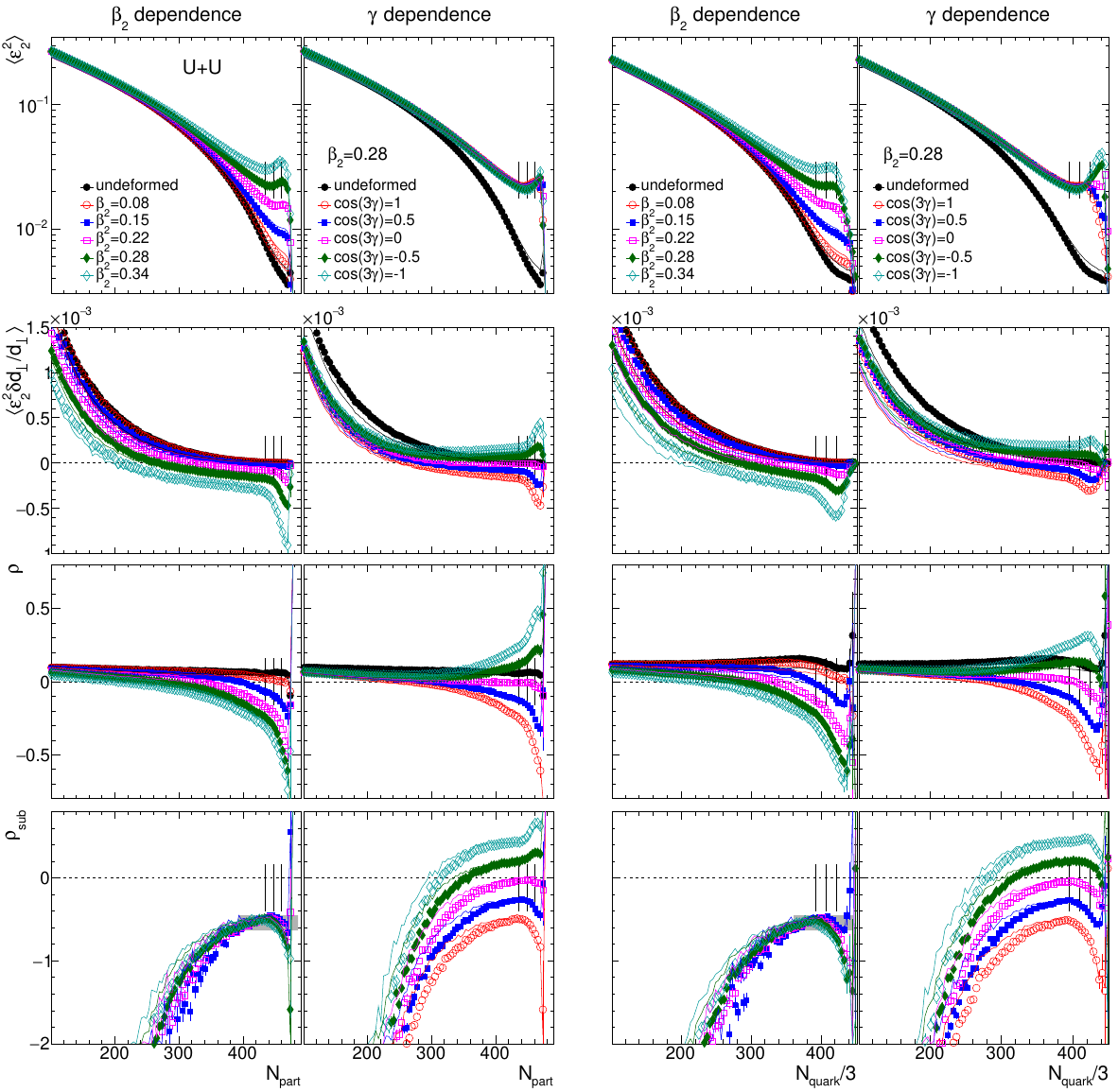}
\end{center}
\caption{\label{fig:app3} $\npart$ (left two columns) and $\nqp$ (right two columns) dependencies of various observables related to $\varepsilon_2$ in U+U collisions compared between different $\beta_2$ (1st and 3rd columns) and different $\gamma$ with $\beta_2=0.28$ (2nd and 4th columns), calculated with nucleons (symbols) and quarks (lines). The vertical lines in each panel correspond to locations for 0.2\%, 1\% and 2\% centralities.}
\end{figure}

\begin{figure}[h!]
\begin{center}
\includegraphics[width=0.99\linewidth]{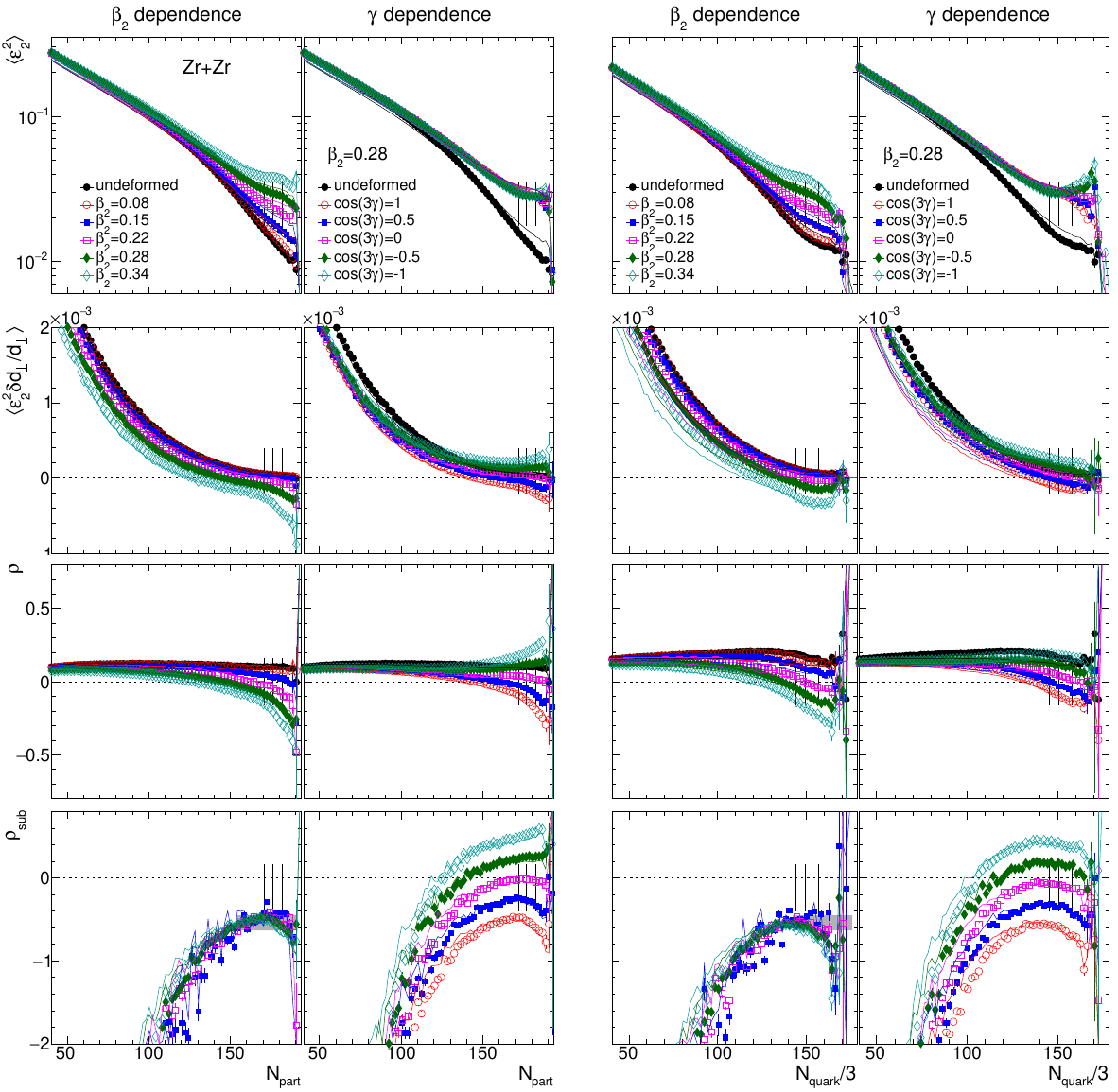}
\end{center}
\caption{\label{fig:app4}  Same as Fig.~\ref{fig:app3} but for Zr+Zr collisions.}
\end{figure}

\begin{figure}[h!]
\begin{center}
\includegraphics[width=0.99\linewidth]{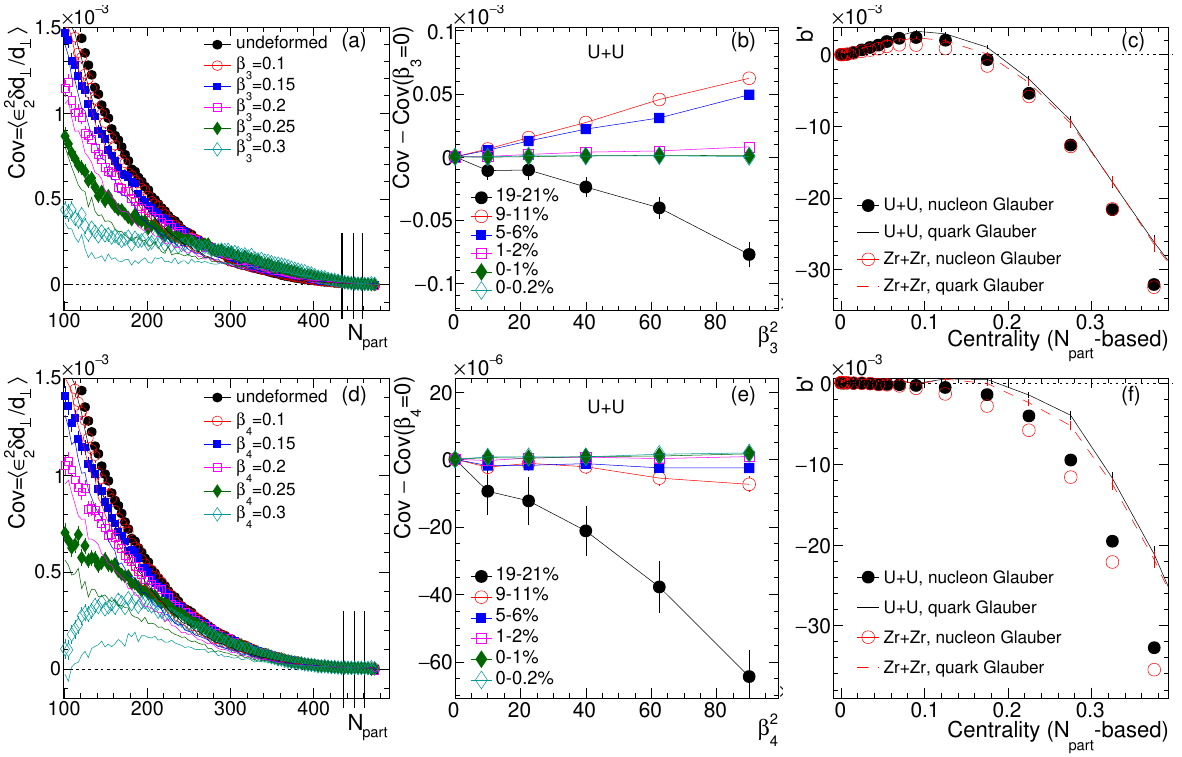}
\end{center}
\caption{\label{fig:app6} The $\lr{\varepsilon_2^2\delta d_{\perp}/d_{\perp}}$ for several values of $\beta_3$ (top row) and $\beta_4$ (bottom row) as a function of $\npart$ (left column) or $\beta_n^2$ (middle column) in U+U collisions. The latter dependencies can be described by a simple $a'+b'\beta_n^2$ function. The right column summarizes the slope $b'$ obtained from the middle panels as a function of centrality in U+U (black) and Zr+Zr (red) systems. }
\end{figure}

\begin{figure}[h!]
\begin{center}
\includegraphics[width=0.99\linewidth]{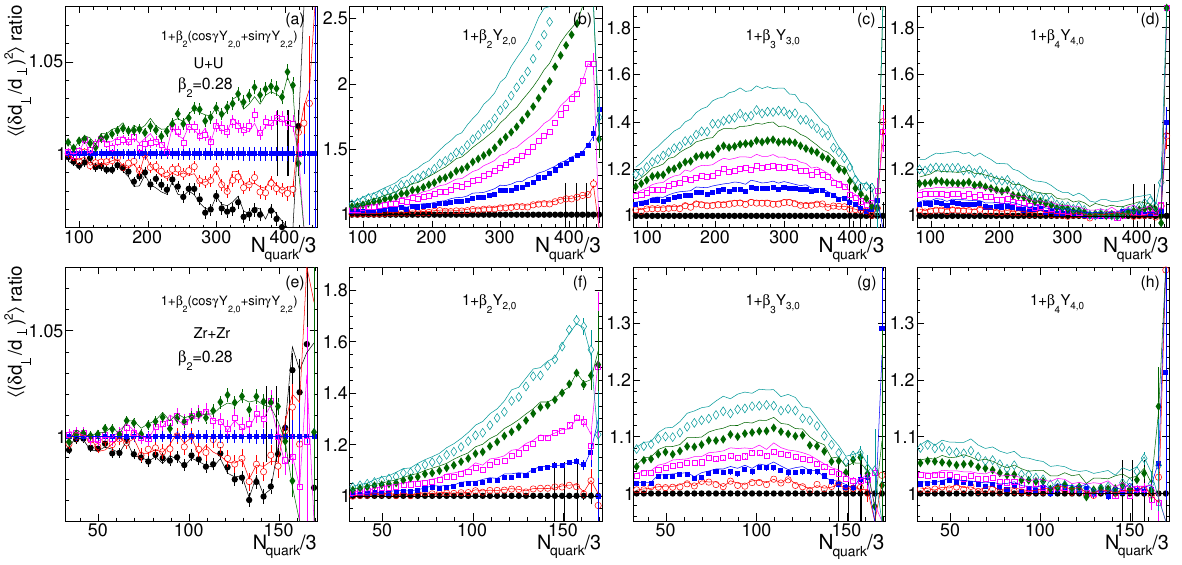}
\end{center}
\caption{\label{fig:app5} Same as Fig.~\ref{fig:5} in the main text, but event average is based on $\nqp$ and plotted as a function of $\nqp$.}
\end{figure}

This appendix shows comprehensive centrality dependence of various observables for different values of $\beta_2$ and $\gamma$ used in the paper in U+U and Zr+Zr collisions. The full set of observables for the cumulants of $d_{\perp}$ are shown in Fig.~\ref{fig:app1} for U+U and Fig.~\ref{fig:app2} for Zr+Zr, respectively. Similarly, information for $\lr{\varepsilon_2^2}$ and correlation between $\varepsilon_2$ and $d_{\perp}$ are shown in Figs.~\ref{fig:app3} and ~\ref{fig:app4}. Most importantly these plots show the results obtained with the $\nqp$-based event averaging. See also Figs.~\ref{fig:app6} and \ref{fig:app5}.


\clearpage
\bibliography{deform3.bib}

\end{document}